\documentclass[journal=jctc,manuscript=article]{achemso}
\usepackage{chemformula} 
\ProvideTextCommand{\DJ}{OT1}{\leavevmode\raisebox{-.5ex}{\makebox[0pt][l]{\hskip-.07em\accent"16\hss}}D}
\usepackage{adjustbox}
\usepackage{caption}
\usepackage{subcaption}
\usepackage{color}
\usepackage{hyperref}
\usepackage{booktabs}
\usepackage{multirow}

\usepackage[flushleft]{threeparttable}
\usepackage{booktabs,caption}
\usepackage{amsmath}

\author{Sree Harsha Bhimineni}
\affiliation[NIT Warangal]
{Chemical Engineering Department, National Institute of Technology, Warangal, Telangana, India}
\altaffiliation{Contributed equally to this work}
\author{Tianhang Zhou}
\affiliation[CUPB]{State Key Laboratory of Heavy Oil Processing, College of Carbon Neutrality Future Technology, China University of Petroleum (Beijing), Beijing 102249, China}
\affiliation[TU-Darmstadt]{Eduard-Zintl-Institut f\"{u}r Anorganische und Physikalische Chemie, Technische Universit\"{a}t Darmstadt, Alarich-Weiss-Str. 8, 64287 Darmstadt, Germany}
\altaffiliation{Contributed equally to this work}
\email{zhouth@cup.edu.cn}
\author{Saeed Mahmoodpour}
\affiliation[TU-Darmstadt]{Technische Universität Darmstadt, Institute of Applied Geosciences, Group of Geothermal Science and Technology, Schnittspahnstrasse 9, 64287 Darmstadt, Germany}
\author{Mrityunjay Singh}
\affiliation[TU-Darmstadt]{Technische Universität Darmstadt, Institute of Applied Geosciences, Group of Geothermal Science and Technology, Schnittspahnstrasse 9, 64287 Darmstadt, Germany}
\author{Wei Li}
\affiliation[TU-Darmstadt]{Department of Materials and Earth Sciences, Technische Universität Darmstadt, Darmstadt 64287, Germany}
\author{Saientan Bag}
\affiliation[TU-Darmstadt]{Eduard-Zintl-Institut f\"{u}r Anorganische und Physikalische Chemie, Technische Universit\"{a}t Darmstadt, Alarich-Weiss-Str. 8, 64287 Darmstadt, Germany}
\author{Ingo Sass}
\affiliation[TU-Darmstadt]{Technische Universität Darmstadt, Institute of Applied Geosciences, Group of Geothermal Science and Technology, Schnittspahnstrasse 9, 64287 Darmstadt, Germany}
\author{Florian M\"{u}ller-Plathe}
\affiliation[TU-Darmstadt]
{Eduard-Zintl-Institut f\"{u}r Anorganische und Physikalische Chemie, Technische Universit\"{a}t Darmstadt, Alarich-Weiss-Str. 8, 64287 Darmstadt, Germany}

\title[An \textsf{achemso} demo]
  {Machine-Learning-Assisted Investigation of the Diffusion of Hydrogen in Brine by Performing Molecular Dynamics Simulation}

\abbreviations{IR,NMR,UV}

\DeclareUnicodeCharacter{2212}{-}

\begin{document}
\pagebreak
\begin{abstract} 
Deep saline aquifers are one of the best options for large-scale and long-term hydrogen storage. Predicting the diffusion coefficient of hydrogen molecules at the conditions of saline aquifers is critical for the modelling of hydrogen storage. 
The diffusion coefficient of hydrogen molecules in chloride brine with different cations ($\mathrm{Na}^+$, $\mathrm{K}^+$, $\mathrm{Ca}^{2+}$) containing up to 5 $\mathrm{mol/kg_{H_2O}}$ concentration is numerically investigated using molecular dynamics (MD) simulation. A wide range of pressure (1-218 atm) and temperature (298-648 K) conditions is applied to cover the realistic operational conditions of the aquifers. We find that the temperature, pressure and properties of ions (compositions and concentrations) affect the hydrogen diffusion coefficient. An Arrhenius behavior of the effect of temperature on the diffusion coefficient has been observed with the temperature independent parameters fitted using the ion concentration under constant pressure. However, it is noted that the pressure strongly affects the diffusive behavior of hydrogen at the high temperature ($\geq$ 400 K) regime, indicating the inaccuracy of the Arrhenius model.
Hence, we combine the obtained MD results with four models of machine learning (ML), including linear regression (LR), random forest (RF), extra tree (ET) and gradient boosting (GB) to provide effective predictions on the hydrogen diffusion. The resultant combination of GB model with MD data predicts the diffusion of hydrogen more effectively as compared to the Arrhenius model and other ML models. Moreover, a \textit{post hoc} analysis (feature importance rank) has been performed to extract the correlation between physical descriptors and simulation results from ML models. Our work provides a promising route for a quick and cost-effective diffusion coefficient determination for multiple and complex brine solutions with a wide range of temperature, pressure and ion concentration by the combination of MD simulations and ML techniques. 
\end{abstract}

\pagebreak

\section{Introduction}{
The development of wind and solar energy is the most promising of all types of renewable energy. However, wind and solar power are difficult to predict accurately due to their intermittency, which inevitably leads to an imbalance between peak demand and peak production \cite{RN1874}. On the other hand, these fluctuations also cause technical challenges such as congestion in transmission lines and disturbances in the power markets. The key to solving intermittent problems is to establish a large-scale energy storage system across seasonal time scales.
 

Hydrogen, as an alternative energy carrier, has characteristics of high specific energy capacity (120 MJ/kg) \cite{RN1887} and high conversion efficiency, and has proven to be one of the lowest cost power storage options \cite{RN1893}.
A recent report for the International Hydrogen Energy Council has stated that by 2050, hydrogen energy demand will account for more than 18 $\%$ of global final energy consumption \cite{RN1894}. Hydrogen can be produced using the concept of “Power to Gas” \cite{RN1895, RN1896, RN1897} by means of electrolysis, utilizing a large amount of surplus renewable electricity generated by renewable solar and wind energy, which helps alleviate the main shortcomings of renewable energy generation: intermittency, seasonality and geographic limitations \cite{RN1884, RN1889, RN1891}. The stored hydrogen can be used as fuel for electricity generation to make up for the demand in the low power generation period (“Gas to Power”) by means of fuel cells \cite{RN1898, RN1899}. It can also be converted into liquid fuel or used as a substance in various industries\cite{ramachandran1998overview}. However, the large-scale hydrogen storage is very challenging because of its very low density (0.08988 g/L) \cite{RN1900} at standard conditions. Furthermore, there are some challenges associated with the storage of materials.

In 2019, the International Energy Agency reported that deep saline reservoirs located at 500 to 2000 m below the Earth’s surface are one of the best options for large-scale and long-term hydrogen energy storage \cite{RN1901}. Saline aquifers are porous media with substantial potential storage capacity (hundreds of $km^{3}$) and are found all over the world \cite{RN1903}. In a saline aquifer, compressed hydrogen replaces the saline water in the micron-sized pores\cite{RN1904, RN1902}, which is very beneficial for continuous flow injection and extraction of hydrogen. The storage capacity of hydrogen is affected by many factors, such as the reservoir’s porosity, temperature, and pressure. In addition, loss of hydrogen is also possible because of the reactions with microorganisms, rocks and fluids. The hydrogen loss caused by the diffusion of hydrogen in rocks and fluids can be estimated by the mass transfer across the boundary layer \cite{Zhao2021}, which is a function of the molecular diffusion coefficient and the concentration gradient of components \cite{RN1905, RN1890}. Hence, it is necessary to study the diffusion coefficient of hydrogen at the conditions of the aquifer.

As the conditions in the aquifer are extreme, the experimental study of diffusion coefficient is challenging and expensive. 
Molecular dynamics (MD) simulations are an attractive alternative. 
The different diffusion coefficients (Self-, Maxwell-Stefan (MS), and Fick) have been intensively evaluated for pure, binary, or multicomponent systems  \cite{krishna2005darken, wheeler2004molecular,liu2011multicomponent,liu2011fick} at different conditions by MD simulation. Diffusion coefficients of various gases such as $\mathrm{CH_4}$, $\mathrm{CO_2}$, and $\mathrm{N_2}$ in water/brine have been measured by MD simulations  \cite{moradi2020prediction,moultos2014atomistic,omrani2021molecular,sharma2014temperature}. A study by Moultos \textit{et al.} in 2014 examined various combinations of $\mathrm{CO_2}$ and water force fields to predict $\mathrm{CO_2}$ diffusion in water for a wide range of pressures and temperatures \cite{moultos2014atomistic}. More recently, Zhao \textit{et al.} proposed a correlation for predicting the diffusion of $\mathrm{H_2}$, $\mathrm{CH_4}$, $\mathrm{CO}$, $\mathrm{O_2}$, and $\mathrm{CO_2}$ in water near the critical point, using MD simulations \cite{Zhao2021}. Omrani \textit{et al.} \cite{omrani2021molecular} used MD simulations to predict $\mathrm{CO_2}$ diffusion coefficient in water/brine for a wide spectrum of pressure and temperature. They also considered multicomponent systems of $\mathrm{CO_2}$-$\mathrm{SO_2}$-water and $\mathrm{CO_2}$-$\mathrm{N_2}$-water to demonstrate the effects of impurities on $\mathrm{CO_2}$ diffusion coefficient in water with implication to $\mathrm{CO_2}$ sequestration  \cite{omrani2021diffusion}. Recently, Bartolomeu and Franco studied the thermophysical properties of supercritical $\mathrm{H_2}$ including the self-diffusion coefficient by MD simulation \cite{bartolomeu2020thermophysical}.  Overall, MD simulation is a valuable and accurate method for calculating diffusion coefficients under different conditions, especially when experiments cannot be carried out.

On the other hand, many studies have indicated the possibility of coupling machine learning (ML) with MD simulation. Allers \textit{et al.} \cite{Allers2020} have integrated ML models 
with the MD database on the prediction of the self diffusion coefficients of Lennard-Jones fluids. They found that the artificial neural net regression models provided superior prediction
of diffusion in comparison to the existing empirical relationships. Liang \textit{et al.} \cite{Liang2021} have obtained the accurate thermophysical properties of $\mathrm{MgCl_2}$ - KCl eutectics such as density, viscosity, diffusion coefficients, and heat capacity from the MD simulation by training the ML-based deep potential. Kirch \textit{et al.} \cite{Kirch2020} have combined MD results of the brine-oil interfacial tension with ML algorithms to obtain a predictive model and estimate the effect of different parameters. 
Recently, we have combined MD simulations and a genetic algorithm to design the sequence-specific block copolymers with a high thermal conductivity\cite{TZhou2021}. 

In this work, we have performed MD simulations to qualitatively study and generate a compatible data set of diffusion coefficients of hydrogen into brine. The results obtained from the MD simulations have been validated by comparing them with the experimental data. The quantitative estimation of the diffusion coefficient has been studied by using two formulations. Firstly, we have investigated the influence of temperature on the diffusion coefficient by the Arrhenius equation. The effect of ion concentration has been incorporated into the temperature independent parameters of the equation. Secondly, we have adapted different ML models trained by the MD data set to investigate the complex effect of temperature, pressure and ion concentration in this work. Moreover, the gradient boosting model has been used to estimate the importance rank of different parameters on the diffusion coefficient of hydrogen molecules.
}

\section{Methodology}
\subsection{MD Simulation}
We use the TIP4P/2005 model\cite{Abascal2005} for $\mathrm{H_2O}$ because of its applicability and accuracy in wide temperature and pressure ranges\cite{Moultos2014}. A two-site model\cite{Yang2005} fitted by the experimental $PVT$ curve of bulk $\mathrm{H_2}$ and validated by the diffusion data is then used for $\mathrm{H_2}$. For the chloride ($\mathrm{Cl^-}$) ions, we use the OPLS model\cite{Chandrasekhar1984}. For cations, the OPLS model\cite{Chandrasekhar1984} ($\mathrm{Na^+}$ and $\mathrm{K^+}$) and Aqvist model\cite{Aqvist1990} ($\mathrm{Ca^{2+}}$) are used since they have been validated against densities and osmotic pressures of aqueous solutions over a large temperature and salinity range \cite{Neyt2013,Tsai2015}. The potential function, $U_{ij}$, used to describe the non-bonded interactions between two particle $i$ and $j$ with a Lenard-Jones potential, and an electrostatic term (expressed by a Coulomb potential) is: 

\begin{equation}
U_{ij} = 4\varepsilon_{ij}\left[\left(\frac{\sigma_{ij}}{r_{ij}}\right)^{12}-\left(\frac{\sigma_{ij}}{r_{ij}}\right)^{6}\right] + \frac{q_iq_j}{4\pi\varepsilon_0 r_{ij}} 
\label{eq:LJ}
\end{equation}
where \(\varepsilon_{ij}\) and \(\sigma_{ij}\) are the Lennard-Jones interaction parameters between the atoms $i$ and $j$; $r_{ij}$ is the distance between them; $q_i$ and $q_j$ are their charges, and \(\varepsilon_0\) is the vacuum permittivity. 
{The force field parameters of the models used in this study are tabulated in Table \ref{table:ffp}.

\begin{table}[H]
\centering
\caption{Force field parameters used for $\mathrm{H_2O}$ with TIP4P/2005 model\cite{Abascal2005}, $\mathrm{H_2}$ with Yang's model\cite{Yang2005}, $\mathrm{Cl^-}$ with OPLS model\cite{Chandrasekhar1984}, $\mathrm{Na^+}$ and $\mathrm{K^+}$ with OPLS model\cite{Chandrasekhar1984} and $\mathrm{Ca^{2+}}$ with Aqvist model\cite{Aqvist1990} in this work.}
\label{table:ffp}
\begin{threeparttable}
    \begin{tabular}{c|c|c|c|c}
    \toprule
         Molecule & Force center or charge & $\mathrm{\varepsilon}$ (kcal/mol) & $\mathrm{\sigma}$ ({\AA}) & q (e) \\
         \midrule
         \multirow{3}{*}{$\mathrm{H_2O}$} & O & 0.185200	& 3.1589 & -1.1128 \\
         & $\mathrm{H1}$ & 0 & 0 & 0.5564 \\
         & $\mathrm{H2}$ & 0 & 0 & 0.5564 \\         
         \multirow{2}{*}{$\mathrm{H_2}$} & $\mathrm{H1}$ & 0.019890 & 2.7200 & 0 \\
         & $\mathrm{H2}$ & 0.019890 & 2.7200 & 0 \\
         $\mathrm{K^+}$ & $\mathrm{K}$ & 0.000328 & 4.9300 & 1 \\ 
         $\mathrm{Na^+}$ & $\mathrm{Na}$ & 1.607040 & 1.8974 & 1 \\
        $\mathrm{Ca^{2+}}$ & $\mathrm{Ca}$ & 0.449702 & 2.4120 & 2 \\
         $\mathrm{Cl^-}$ & $\mathrm{Cl}$ & 0.117840 & 4.4172 & -1 \\
    \bottomrule
    \end{tabular}
    \end{threeparttable}
\end{table}
}
The Lennard-Jones interaction parameters between unlike atoms \((i \neq j)\) are calculated using the Lorentz-Berthelot mixing rule\cite{Kong1973}:
\begin{equation}
    \varepsilon_{ij} = \sqrt{\varepsilon_i\varepsilon_j}
    \label{eq:epsilon}
\end{equation}
\begin{equation}
    \sigma_{ij} = \frac{\sigma_i + \sigma_j}{2}
    \label{eq:sigma}
\end{equation}

MD simulations are performed in the Large-scale Atomic/Molecular Massively Parallel Simulator (LAMMPS). \cite{Plimpton1997} 
Three types of systems, namely single cation systems ($\mathrm{K^+}$, $\mathrm{Na^+}$, and $\mathrm{Ca^{2+}}$), binary cation systems ($\mathrm{K^+}$-$\mathrm{Na^+}$, $\mathrm{K^+}$-$\mathrm{Ca^{2+}}$ and $\mathrm{Na^+}$-$\mathrm{Ca^{2+}}$), and ternary cation systems ($\mathrm{Na^+}$-$\mathrm{K^+}$-$\mathrm{Ca^{2+}}$) are simulated. Examples of the compositions are summarized in Table \ref{concentration}. The number of ions is chosen according to the desired molality. For example, 216 particles of $\mathrm{Na^+}$ and $\mathrm{Cl^-}$ represent a molality of 3 $\mathrm{mol/kg_{H_{2}O}}$ in 4000 $\mathrm{H_2O}$ molecules. To cover a wide range of the brine composition, literature is examined and 5 $\mathrm{mol/kg_{H_{2}O}}$ is selected as the upper limit for the salinity in this study \cite{wellman2003evaluation, gaus2005reactive, xu2006toughreact, bacon2009reactive, azin2013measurement, mohamed2013effect, wang2016study, vu2017changes, shi2018measurement, mahmoodpour2019effect}. The total charge is zero in all systems.
The cut-off distance for the Lennard-Jones potential and the real-space Coulombic cut-off distance have been used as 12 \r{A} and 8.5 \r{A}, respectively. It should be noted that the long-range part of the electrostatic forces is evaluated using the particle-particle particle-mesh (PPPM) solver\cite{RogerW.Hockney;JamesW.Eastwood1988}. The bonds and angles of $\mathrm{H_2}$ and $\mathrm{H_2O}$ molecules are fixed using the SHAKE algorithm\cite{Ryckaert1977}. For all the simulations, a 0.4 ns preliminary run with \(T_0\) = 300 K and \(P_0\) = 1 atm have been performed, followed by a 1.1 ns equilibration in the isothermal-isobaric (NPT) ensemble using the Nosé-Hoover thermostat and barostat with a temperature damping parameter of 100 fs and pressure damping parameter of 1000 fs. Three production runs with each of 2 ns are run in the microcanonical (NVE) ensemble, and the simulations are sampled every 10 ps to perform the analysis. The time step used for all the runs is 1 fs.
\begin{table}[H]
\centering
\caption{Clarification of brine compositions of the systems simulated in this work.}
\begin{threeparttable}
\begin{tabular}{c c | c c c c| c}
\toprule
{} & Systems & $\mathrm{c_{Na^+}}$* & $\mathrm{c_{Ca^{2+}}}$* & $\mathrm{c_{K^+}}$* & $\mathrm{c_{Cl^-}}$* & c* \\
\midrule
    {} & Single & 3.0 &  0.0  &  0.0 &  3.0  &  3.0 \\
    {} & Single & 0.0 &  3.0  &  0.0 &  6.0  &  3.0 \\
    {} & Single & 0.0 &  0.0  &  3.0 &  3.0  &  3.0 \\
    {} & Binary & 1.5 &  1.5  &  0.0 &  4.5  &  3.0 \\
    {} & Binary & 0.0 &  1.5  &  1.5 &  4.5  &  3.0 \\
    {} & Binary & 1.5 &  0.0  &  1.5 &  3.0  &  3.0 \\
    {} & Ternary & 1.0 &  1.0  &  1.0 &  4.0  &  3.0 \\
 \bottomrule
\end{tabular}
\begin{tablenotes}
\small
\item *Concentrations are in $\mathrm{mol/{kg_{H_2O}}}$.
\item $\mathrm{c_{Na^+}}$: Concentration of $\mathrm{Na^+}$ ions; $\mathrm{c_{Ca^{2+}}}$: Concentration of $\mathrm{Ca^{2+}}$ ions; $\mathrm{c_{K^+}}$: Concentration of $\mathrm{K^+}$ ions and $\mathrm{c}$: Total cation concentration.
\end{tablenotes}
\end{threeparttable}
\label{concentration}
\end{table}


The diffusion coefficient is obtained from the Einstein relation of the mean squared displacement (MSD) with diffusion time\cite{Einstein1905} as follows:

\begin{equation}
    D = \lim_{t \to \infty}\frac{\langle [r_i(t)-r_i(0)]^2 \rangle}{6t}
    \label{eq:D}
\end{equation}
where \(r_i(t)\) is the position vector of $\mathrm{H_2}$ at time $t$, and \(\langle\)   \(\rangle\) indicates the average over all $\mathrm{H_2}$ molecules and time origins. The viscosity is determined according to the Green-Kubo formula \cite{green1954markoff, kubo1957statistical}: 

\begin{equation}
    \eta = \frac{V}{k_BT}\int_{0}^{\infty}\langle P_{\alpha\beta}(t_0)P_{\alpha\beta}(t_0+t)\rangle dt
    \label{eqn:vis}
\end{equation}
where $k_B$ is the Boltzmann constant, $V$ is the volume of the simulation box, and the $P_{\alpha \beta}$ are the off-diagonal components of the stress tensor.

\subsection{Machine Learning}

\begin{figure}[H]
	\centering
	  \includegraphics[scale=0.65]{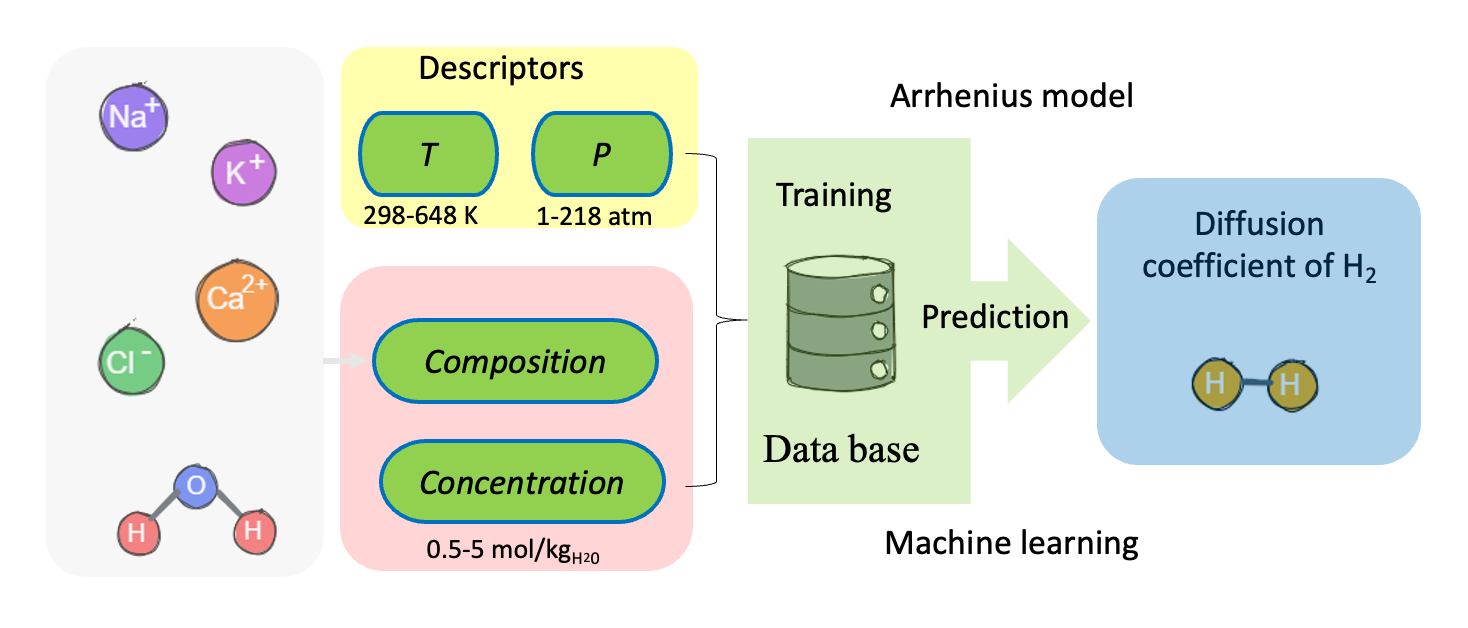}
	  \caption{The diffusion coefficient of hydrogen is predicted through Arrhenius-fitting and machine learning methods in this work. The temperature, pressure, concentration and composition of ions are input features.}
	  \label{fig:system}
\end{figure}

By the use of the scikit-learn package \cite{pedregosa2011scikit}, four different machine learning models have been implemented on the simulated MD data to give predictions of the diffusion coefficient (Figure \ref{fig:system}). They are (i) linear regression (LR), (ii) random forest (RF), (iii) extra trees (ET) and (iv) gradient boosting (GB). Hyperparameters which are tunable parameters used to control the learning process are tuned to optimize the models. 
The data set contains 5 features: temperature (T), pressure (P), and three cation concentrations ($\mathrm{c_{Na^+}}$, $\mathrm{c_{Ca^{2+}}}$ and $\mathrm{c_{K^+}}$) which are used to predict the diffusion coefficient. Details of these models have been summarized in the Supporting Information.

The mean absolute error (MAE) and root mean squared error (RMSE) are evaluated to compare the accuracy of the models as follows:
\begin{equation}
e_{MAE} = \frac{\sum_{i=1}^{m} |y_i - y_{pi}|}{m}
\label{eq6}
\end{equation}

\begin{equation}
e_{RMSE} = \sqrt{\frac{\sum_{i=1}^{m} (y_i - y_{pi})^2}{m}}
\label{eq7}
\end{equation}

where $y_i$ and \textit{y}\textsubscript{\textit{pi}} represent the actual and the predicted values of the diffusion coefficient, respectively. The coefficient of determination ($R^2$ score), another metric to compare the accuracy, is given by:

\begin{equation}
R^2 = 1-\frac{SS_{res}}{SS_{tot}}
\label{eq8}
\end{equation}

where \textit{SS}\textsubscript{res} is the sum of squared residuals given by $\sum_{i=1}^{m} (y_i-y_{pi})^2$ and \textit{SS}\textsubscript{tot} is the total sum of squares given by $\sum_{i=1}^{m} (y_i-y_{av})^2$ where ${y_{av}}$ is the average of the actual values of diffusion coefficients. The best possible score of \textit{R}$^{2}$ is 1.0, under the circumstance that all data is predicted correctly. A model that gets a zero score of \textit{R}$^{2}$ predicts the expected value of the data set without considering the input features. 
Besides the $R^2$ score calculated using the testing data for the comparison of the accuracy of ML models, the mean cross-validated $R^2$ score (MCRS)\cite{CV} is calculated over 10 cross-validation splits of the data with Eq. \ref{eq8} to tune the hyperparameters of the ML models. 

Moreover, the GB model has been used to acquire the feature importance rank (FIR) in this work. Feature importance is the normalized measure of the frequency of an individual feature to be used in the training process. For a higher FIR, the specific feature is used more frequently to split the nodes. As a result, it provides the information of the correlation of different features with the diffusion coefficient of hydrogen molecules.

\section{Results and Discussions}
\subsection{Validations}
\begin{figure}[H]
	\centering
	  \includegraphics[scale=0.7]{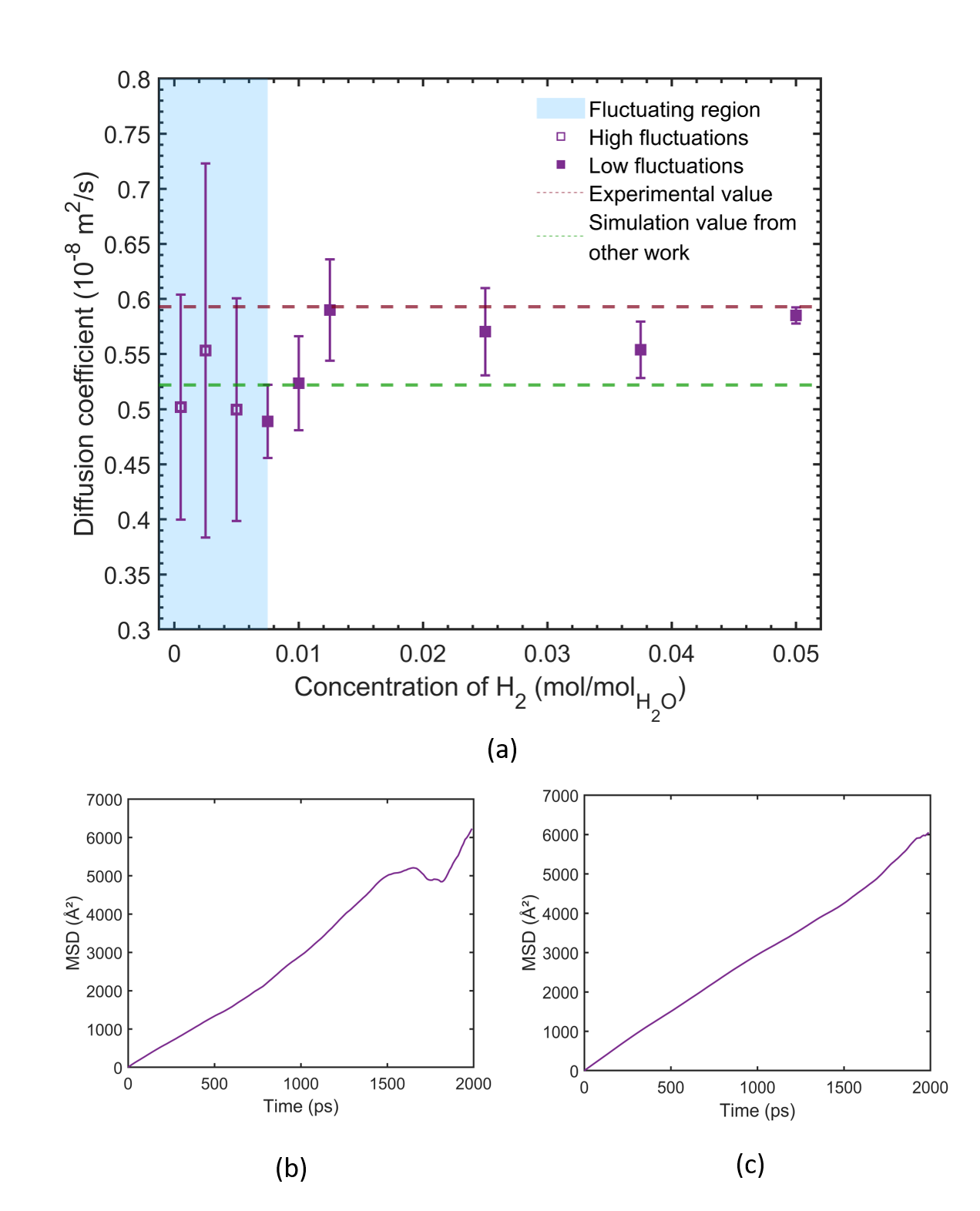}
	  \caption{(a) The variation of the diffusion coefficient of $\mathrm{H_2}$ in pure $\mathrm{H_2O}$ versus the concentration of $\mathrm{H_2}$ at T = 298 K and P = 1 atm. The dashed red line is obtained from Eq. \ref{eq:wilke} using the data provided by Wise \textit{et al.}\cite{Wise1966} and the dashed green line is the result from other molecular dynamics (MD) work by Kallikragas \textit{et al.} \cite{Kallikragas2014}. Mean squared displacement (MSD) versus time for systems with (b) 2 $\mathrm{H_2}$ molecules and (c) 30 $\mathrm{H_2}$ molecules at $T$ = 298 K and $P$ = 1 atm.}
	  \label{fig:validation}
\end{figure}

\begin{figure}[H]
    \centering
    \includegraphics[scale=0.8]{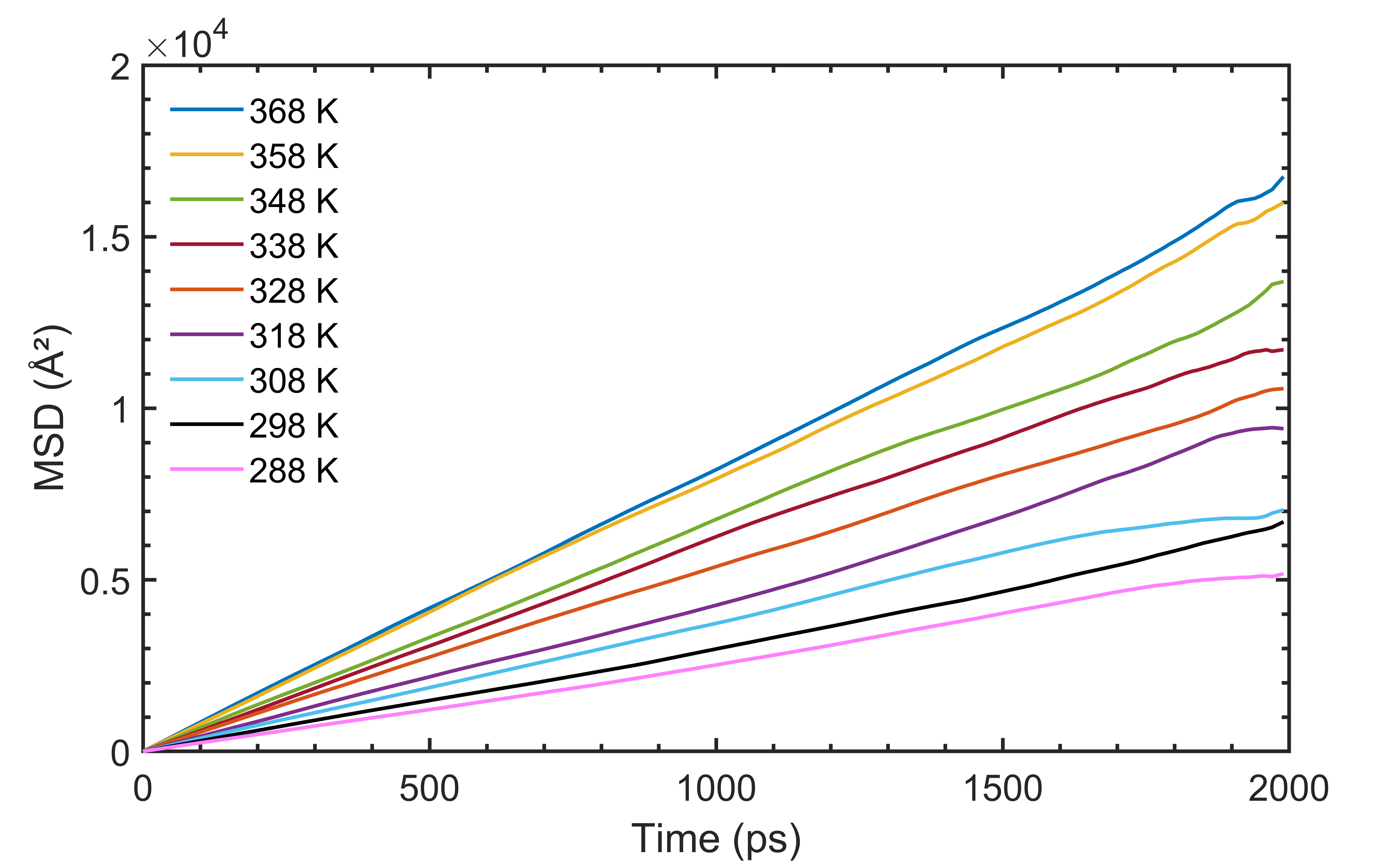}
    \caption{The variation of mean squared displacement (MSD) of $\mathrm{H_2}$ in pure $\mathrm{H_2O}$ versus time with respect to different temperatures (288, 298, 308, 318, 328, 338, 348, 358 and 368 K) at 1 atm.}
    \label{fig:msdvst}
\end{figure}

{
In order to check the effect of $\mathrm{H_2}$ concentration on the diffusion coefficient, we have examined systems with 2, 10, 20, 30, 40, 50, 100, 150 and 200 $\mathrm{H_2}$ molecules with 4000 water molecules (concentration of $\mathrm{H_2}$ $\sim$ 0.0005 $\mathrm{mol/mol_{H_2O}}$ - 0.05 $\mathrm{mol/mol_{H_2O}}$) at 298 K and 1 atm. The diffusion coefficient of $\mathrm{H_2}$ fluctuates with large errors in case of the low solubility (Figure \ref{fig:validation} (a)). This can be explained by the significant deviation of MSD from the linear relationship with poor statistics, as shown in Figure \ref{fig:validation} (b). The deviation is reduced and the diffusion coefficient of $\mathrm{H_2}$ is practically constant within its error bars as the number of $\mathrm{H_2}$ molecules increases beyond 30 (Figure \ref{fig:validation} (c)). Wilke \textit{et al.}\cite{Wilke1955} proposed an empirical equation for the diffusion coefficients of gas in water as: 
\begin{equation}
    D \propto \frac{T}{\mu}
    \label{eq:wilke}
\end{equation}
where T is the temperature and $\mu$ is the viscosity. We have inserted experimental results\cite{Wise1966} into Eq. \ref{eq:wilke} and interpolated the diffusion coefficient as 0.593$\cdot \mathrm{10^{-8}}$ $\mathrm{m^2/s}$ for $\mathrm{H_2}$ molecules at $T$ = 298 K and $P$ = 1 atm. Details are summarized in the Supporting Information Table S1. On the other hand, Kallikragas \textit{et al}.\cite{Kallikragas2014} have examined MD simulations with SPC/E water potential\cite{berendsen1987missing} and the potential devised by Cracknell\cite{cracknell2001molecular} for $\mathrm{H_2}$ to obtain the diffusion coefficient as 0.522$\cdot \mathrm{10^{-8}}$ $\mathrm{m^2/s}$ for $\mathrm{H_2}$ molecules at $T$ = 298 K and $P$ = 1 atm. The diffusion coefficient of 40 $\mathrm{H_2}$ molecules ($D$ = 0.524$\cdot \mathrm{10^{-8}}$ $\mathrm{m^2/s}$) in this work shows good stability (Figure \ref{fig:validation} (a)) and is comparable to both the above experimental and simulation results. Therefore, the number of 40 $\mathrm{H_2}$ molecules (0.01 $\mathrm{mol/mol_{H_2O}}$) has been used in all simulations in the upcoming discussions. We have also examined the variation of MSD curves of 40 $\mathrm{H_2}$ in pure water at (288, 298, 308, 318, 328, 338, 348, 358 and 368) K and 1 atm in Figure \ref{fig:msdvst}. The MSD curves of different temperatures show a comprehensible difference at the end of the simulation. The linearity of all MSD plots versus time proves that the diffusion coefficients of $\mathrm{H_2}$ can be calculated accurately and credibly. Moreover, Wang\cite{Wang1965} has reported the experimental value of self diffusion coefficient of $\mathrm{H_2O}$ as 0.257$\cdot \mathrm{10^{-8}}$ $\mathrm{m^2/s}$ and Holz \textit{et al.}\cite{Holz2000} have reported it as 0.230$\cdot \mathrm{10^{-8}}$ $\mathrm{m^2/s}$ at $T$ = 298 K and $P$ = 1 atm, which are in close agreement with the value obtained in this work which is  0.233$\cdot \mathrm{10^{-8}}$ $\mathrm{m^2/s}$.

}

	
	

	


\begin{figure}[H]
    \centering
    \includegraphics[scale=1.0]{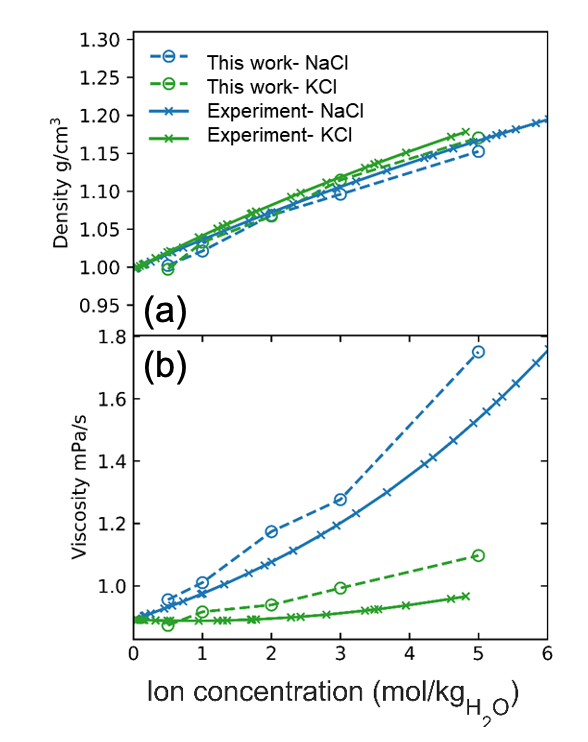}
    \caption{The variation of (a) the density and (b) viscosity of KCl and NaCl solutions versus the ion concentrations at 298 K and 1 atm.\cite{zhang_viscosity_1996}}
    \label{fig:triw}
\end{figure}

We further compare simulated and experimental KCl and NaCl solutions \cite{zhang_viscosity_1996} to validate brine systems (Figure \ref{fig:triw}). Due to the lack in diffusion coefficients for brine systems in the literature, the values of density and viscosity are selected for the comparison. The increasing trend of our simulations are in agreement with the experimental data regarding density (Figure \ref{fig:triw} (a)) and viscosity (Figure \ref{fig:triw} (b)) as the ion concentrations increase.\cite{zhang_viscosity_1996} The magnitude of the deviations in viscosity ((Figure \ref{fig:triw} (b))) could be affected by the complexity of the measurement in the experiment. These observations and discussions confirm that our systems are well-parameterized compared to previous literature.

\subsection{Variation of the $\mathrm{\bf H_2}$ diffusion coefficient}
{
The diffusion coefficients of $\mathrm{H_2}$ in different chloride brines have been calculated at 298 K and 1 atm (Figure {\ref{fig:triw}} (a)) and at  648 K and 218 atm (Figure {\ref{fig:triw}} (b)). The  concentration of chloride brines is varied from 0.5 $\mathrm{mol/kg_{H_{2}O}}$ to 5 $\mathrm{mol/kg_{H_{2}O}}$.
First, it is observed that the diffusion coefficients of $\mathrm{H_2}$ at high temperature and pressure (Figure {\ref{fig:triw}} (b)) are much larger than that at low temperature and pressure (Figure {\ref{fig:triw}} (a)) irrespective of the compositions and molalities of ions. There are previous studies which investigated the effect of temperature and pressure on the diffusion coefficients of gases in pure water and brine \cite{Cadogan2014,Maharajh1972,Chen2018}. The trends of our results are in agreement with the previous studies. 
Second, there is a tendency for the diffusion coefficient to decrease with an increase in the molality for most systems in Figure {\ref{fig:triw}}. However, systems containing single cation $\mathrm{K^+}$ show an increase and then a decreasing trend of the diffusion coefficient versus the molality at high temperature and pressure (Figure \ref{fig:triw} (b)). To understand the mechanism underlying the increased diffusion coefficient, we calculate the intermolecular radial distribution functions (Supporting Information Figure S2). We find that the interaction between $\mathrm{H_2}$ and $\mathrm{K^+}$ becomes weaker as the molality increases, resulting in a lower aberration to the movement of $\mathrm{H_2}$ from $\mathrm{K^+}$ (Supporting Information Figure S2). Then, the the disturbance in the motion of hydrogen molecules could be explained by two compensatory effects, namely, the interaction between $\mathrm{H_2}$ and $\mathrm{K^+}$ and the interaction between $\mathrm{H_2}$ and $\mathrm{H_2O}$. A similar trend has also been observed for water molecules in brine systems containing $\mathrm{K^+}$ \cite{kim_self-diffusion_2012}. Kim \textit{et al.} found that the diffusion coefficient of water molecules with structure-breaking ions (e.g., $\mathrm{K^+}$) increases as the ion concentration increases, whereas the diffusion coefficient of water molecules with structure-making ions (e.g., $\mathrm{Na^+}$ and $\mathrm{Ca^{2+}}$) decreases.\cite{kim_self-diffusion_2012}
However, it is noted that the effect of $\mathrm{Na^+}$ and $\mathrm{Ca^{2+}}$ are more stronger than $\mathrm{K^+}$ in binary and ternary systems, and an overall reduction in diffusion coefficient is observed (Figure {\ref{fig:triw} (b)}). Hence, we focus on the ternary cations systems to evaluate the effect of temperature on the diffusion coefficient of $\mathrm{H_2}$ in the next section. 
\begin{figure}[H]
	  \includegraphics[scale = 0.6]{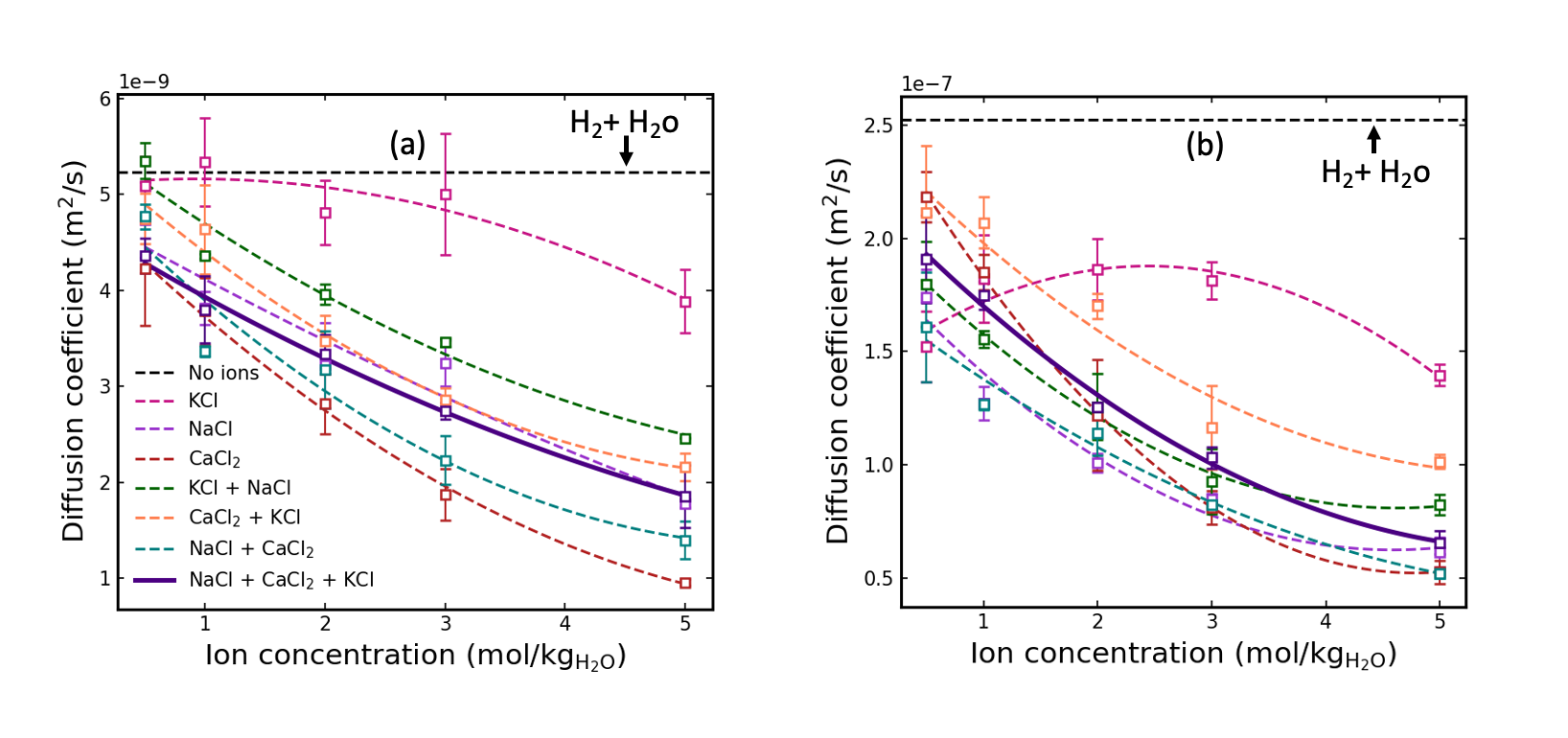}
	  \caption{The variation of the diffusion coefficient of $\mathrm{H_2}$ in saline brine versus the total cation concentration varying from 0.5 $\mathrm{mol/kg_{H_{2}O}}$ to 5 $\mathrm{mol/kg_{H_{2}O}}$ at (a) $T$ = 298 K and $P$ = 1 atm   and (b) $T$ = 648 K and $P$ = 218 atm.}
	  \label{fig:terinaryions}
\end{figure}

}
\subsection{Dependence of the $\mathrm{H_2}$ diffusion coefficient on temperature, pressure and cation concentration}

\begin{figure}[H]
	  \includegraphics[scale = 1]{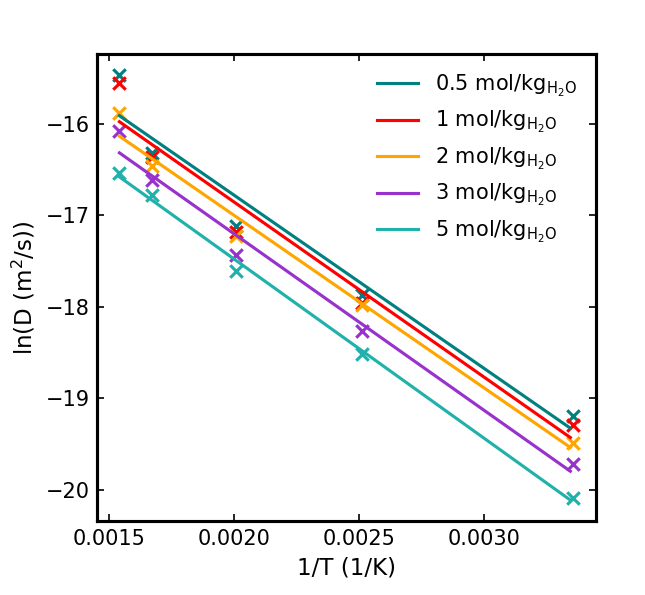}
	  \caption{Diffusion coefficients and the corresponding Arrhenius fits as a function of temperature for ternary cation systems ($\mathrm{Na^+}$-$\mathrm{K^+}$-$\mathrm{Ca^{2+}}$) with total cation concentrations varying from 0.5 $\mathrm{mol/kg_{H_{2}O}}$ to 5 $\mathrm{mol/kg_{H_{2}O}}$ at $P$ = 218 atm.}
	  \label{fig:arrhenius}
\end{figure}

The diffusion coefficient of gases in liquids strongly depends on the temperature.\cite{Wise1966,Cadogan2014,Maharajh1972} 
As a first approximation, the Arrhenius equation\cite{Arrhenius1889} was proposed for the prediction of the temperature dependence of the diffusion coefficient as: 

\begin{equation}
    D = D_0exp\left(\frac{-E_d}{RT}\right)
    \label{eqn: arrhenius}
\end{equation}

where $D_0$ is the diffusion coefficient at infinite temperature, $E_d$ is the diffusion activation energy, $T$ is temperature and $R=8.314$ $\mathrm{J/(mol \cdot K)}$ is the universal gas constant. This formula has been used to describe the diffusion behavior of different gases in water at the constant pressure\cite{Zhao2019,Thapa2013,Jahne1987,Frank1996}. Recently, Chen \textit{et al.} \cite{Chen2018} found that at constant pressure, the temperature dependence of the diffusion coefficient of $\mathrm{CH_4}$ in $\mathrm{NaCl}$, $\mathrm{Ca{Cl_2}}$, $\mathrm{NaBr}$ and $\mathrm{NaI}$ brine can also be predicted by Eq. \ref{eqn: arrhenius}. Similar exponential behavior of the diffusion coefficients of hydrogen in ternary cation systems has been observed in this work (Figure \ref{fig:arrhenius}). 

\begin{figure}[H]
	
	\begin{subfigure}[b]{0.49\linewidth}
		\includegraphics[width=0.925\linewidth]{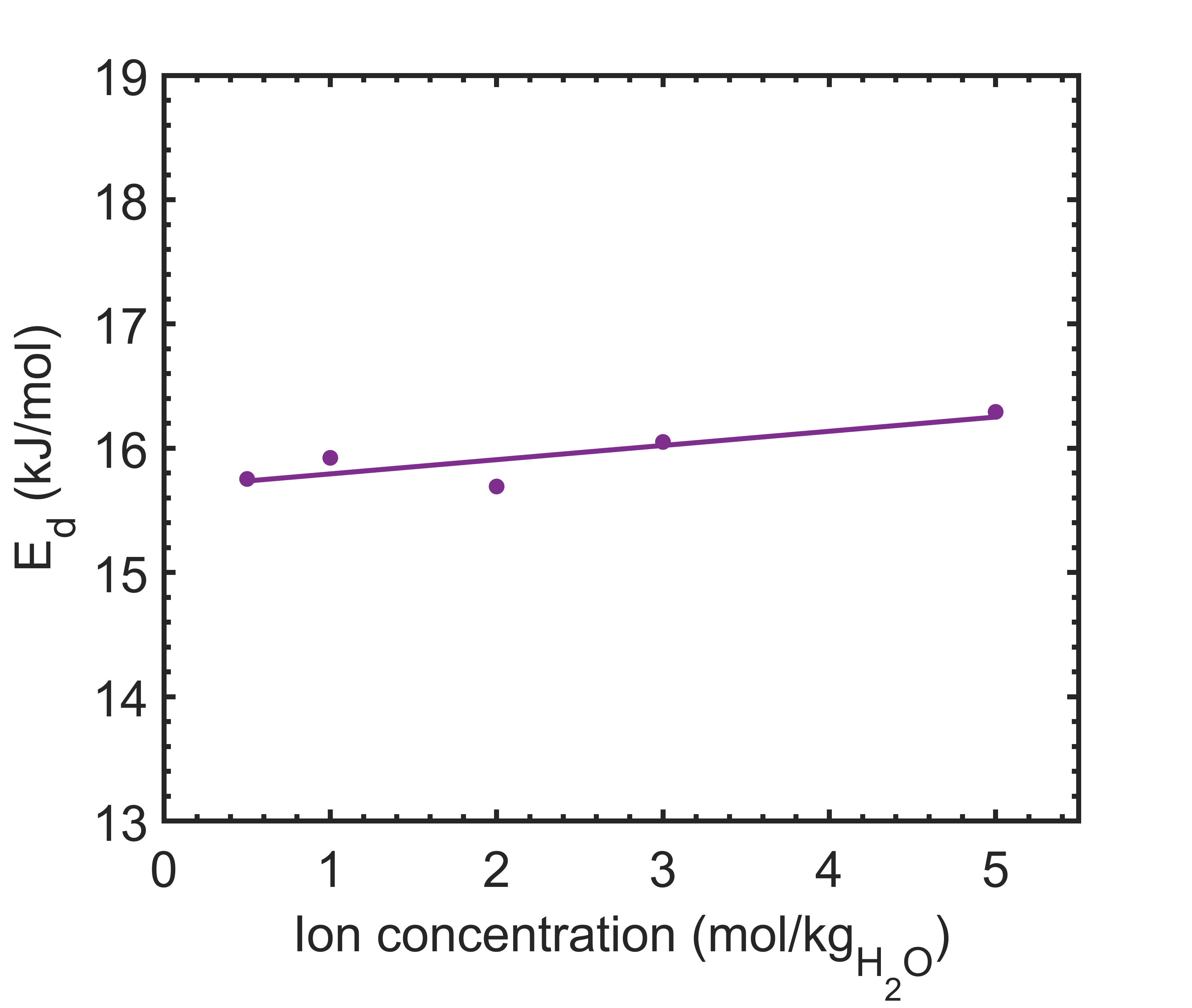}
		\caption{}
		\label{fig: slope}
	\end{subfigure}
	\begin{subfigure}[b]{0.49\linewidth}
		\includegraphics[width=0.99\linewidth]{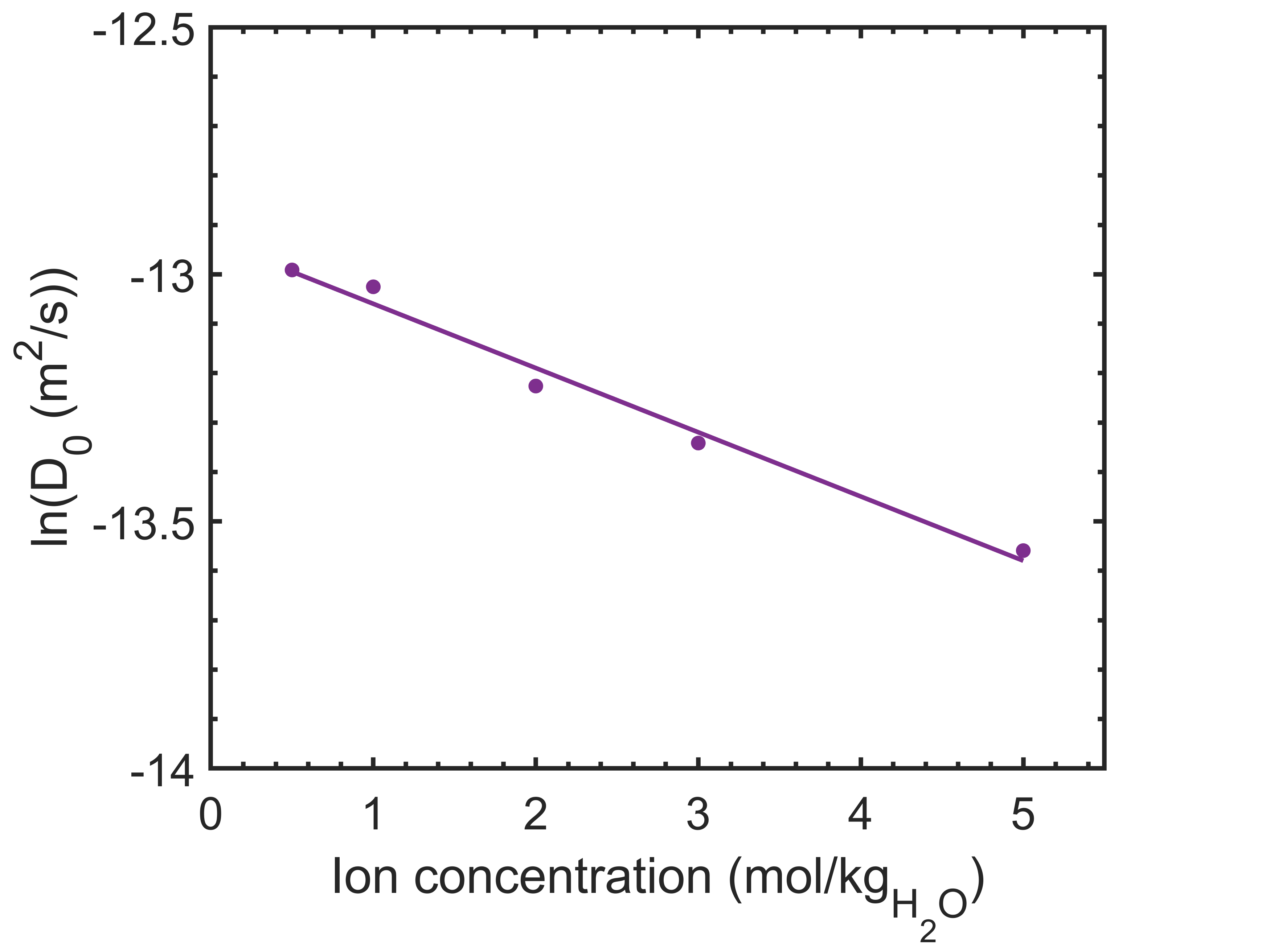}
		\caption{}
		\label{fig: intercept}
  \end{subfigure}

	\caption{Concentration dependence and linear fits of (a) ${E_d}=0.1143c + 15.6728$ and (b) $lnD_0 = -0.1300c - 12.9297$ at $P$ = 218 atm, where $c$ is the total cation concentration.} 
	\label{fig:icf}
	
\end{figure}

A straight line fit has been found for the $lnD$ versus $\frac{1}{T}$ (Eq. \ref{eqn: straightfit}) with different ion concentrations (Figure \ref{fig:arrhenius}) and the $D_0$ and $E_d$ have been extracted.
\begin{equation}
    lnD = \left(\frac{-E_d}{R}\right)\frac{1}{T} + (lnD_0)
    \label{eqn: straightfit}
\end{equation}
Parameters $E_d$ (in kJ/mol) and $lnD_0$ ($D_0$ in $\mathrm{m^2/s}$) are then linearly fitted to the total cation concentrations (Figure \ref{fig:icf} (a) and (b)) as :
\begin{equation}
    {E_d} = 0.1143c + 15.6728
    \label{eq: slope}
\end{equation}
\begin{equation}
    ln(D_0) = -0.1300c - 12.9297
    \label{eq: intercept}
\end{equation}


{
Above discussions on the Arrhenius model of diffusion coefficient is under the constant pressure. We further investigate the effect of pressure on the diffusion coefficient of $\mathrm{H_2}$. At room temperature (298 K), the effect of pressure on the diffusion coefficient is negligible for both low and high ion concentrations (Figure \ref{fig:pe} (a) and (b)), which has also been observed for the diffusion of $\mathrm{CO_2}$ and $\mathrm{N_2}$ in water\cite{Cadogan2014} at room temperature. This implies that the Arrhenius fit Eq.\ref{eqn: straightfit} can be applied regardless of the pressure at room temperature over a pressure range of 1 to 218 atm. However, the diffusion coefficient of $\mathrm{H_2}$ shows a strong dependence on the pressure and cation concentration as the temperature increases from 298 K to 648 K in Figure \ref{fig:pe}. At low ion concentration (Figure \ref{fig:pe} (c)), diffusion coefficient of $\mathrm{H_2}$ decreases with the pressure. The reduction of diffusion coefficient can be explained by the fact that the viscosity of the solution increases as pressure increases\cite{Kestin1981, Mao2009, Martin1983}, which in turn increases the resistance to the motion of the $\mathrm{H_2}$ molecules significantly. At high ion concentrations (Figure \ref{fig:pe} (d)), the diffusion coefficient remains constant with respect to pressure. This non-decreasing trend of diffusion coefficient results from the large number of ions in the system which already confines the motion of the hydrogen molecule due to increased collisions at low pressure\cite{Jaynes1983}. These observations imply that the development of a fitting function which considers the pressure, cation compositions and cation concentrations could be complicated. Zhao \textit{et al}.\cite{Zhao2021} have developed a more exact model for hydrogen in pure water. However, it requires additional information on the density and viscosity of the solution, which require additional efforts in experimenting or simulating for the data. Therefore, new robust models need to be considered. 

	

	

\begin{figure}[H]
    \centering
    \includegraphics[scale = 0.4]{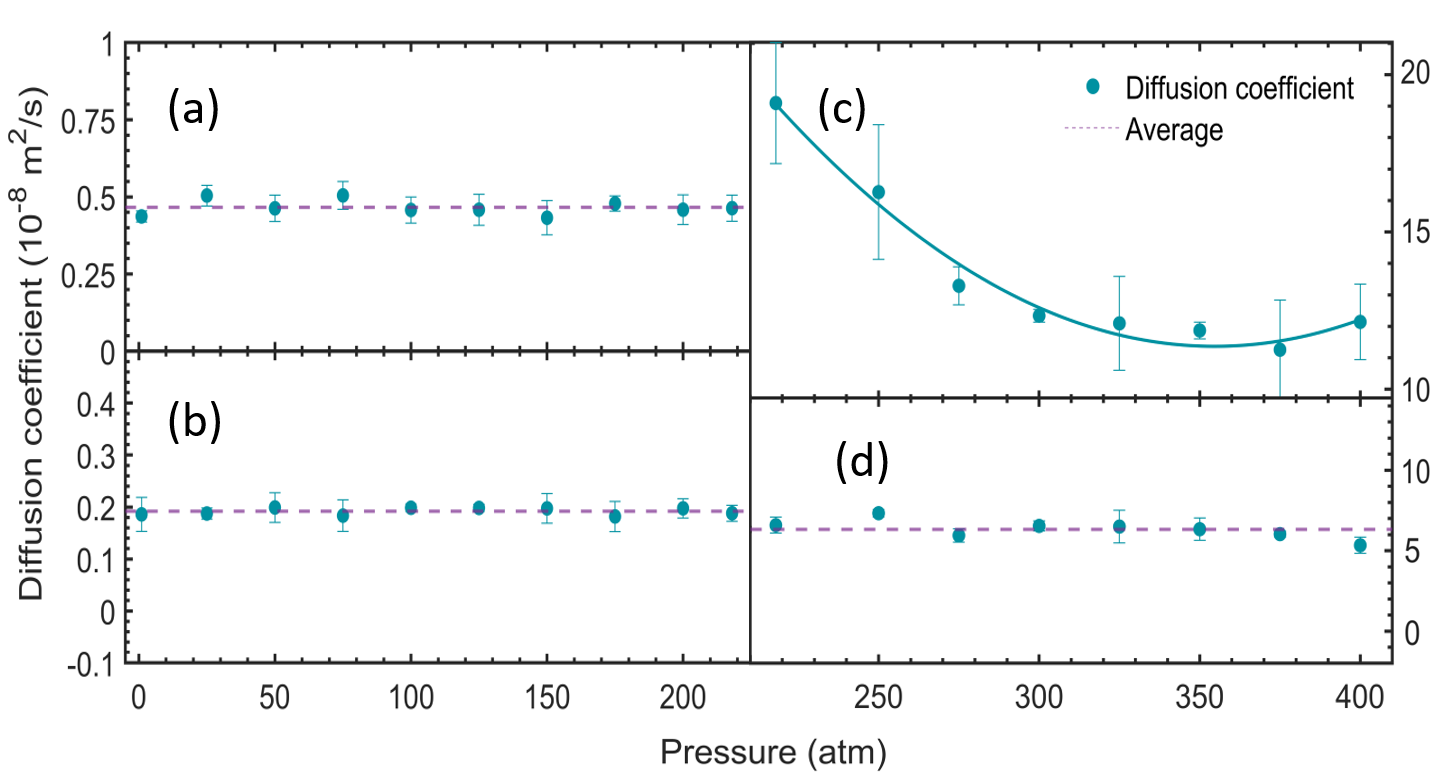}
    \caption{The variation of diffusion coefficient of $\mathrm{H_2}$ in the ternary cation systems ($\mathrm{Na^+}$-$\mathrm{K^+}$-$\mathrm{Ca^{2+}}$) versus pressure at {(a) $T$ = 298 K and low ion concentration $c$ = 0.5 mol/kg$\mathrm{H_2O}$; (b) $T$ = 298 K and high ion concentration $c$ = 5 mol/kg$\mathrm{H_2O}$; (c) $T$ = 648 K and low ion concentration $c$ = 0.5 mol/kg$\mathrm{H_2O}$ and (d) $T$ = 648 K and high ion concentration $c$ = 5 mol/kg$\mathrm{H_2O}$.}}
    \label{fig:pe}
\end{figure}

{
\subsection{Prediction of the diffusion coefficient of $\mathrm{H_2}$ based on machine learning models}  

Here, we use four ML methods to predict the diffusion coefficient concerning temperature, pressure, and ion concentrations. We first identify the proper size of training data, that is, how much data is needed for training to achieve the proper accuracy. Investigation of learning curve has been performed and summarized in the Supporting Information. The prediction accuracy plateaued at about 197 training samples where the data set contained a total of 282 samples. Hence, 70 $\%$ of the data set was selected as training samples. Then, except the simple LR model, the hyperparameters of the other three implemented models were further optimized by the MCRS calculated over ten cross-validation splits to determine the optimal model configuration. The number of estimators (\textit{n}\textsubscript{est}), the depth (\textit{max}\textsubscript{dep}), and the number of features (\textit{max}\textsubscript{feat}) are hyperparameters for the RF, ET, and GB models. Additionally, the influence of the learning rate on the prediction accuracy has also been investigated for the GB model. The detailed optimisation process is summarized in the Supporting Information and obtained hyperparameters are listed in Table~\ref{Tab3.0}.
\begin{table}[H]
  \centering
  \caption{\textbf{Hyperparemeters of the machine learning methods in this work.}}
   \begin{tabular}{ccc}
    \toprule
    Model & Hyperparameter & Value\\
    \midrule
    \multirow{3}{*}{RF} & \textit{n}\textsubscript{est} & 800\\
     & \textit{max}\textsubscript{dep} & 13\\
     & \textit{max}\textsubscript{feat} & 5\\
\hline
    \multirow{3}{*}{ET} & \textit{n}\textsubscript{est} & 100\\
     & \textit{max}\textsubscript{dep} & 10\\
     & \textit{max}\textsubscript{feat} & 5\\
\hline
    \multirow{4}{*}{GB} & \textit{n}\textsubscript{est} & 600\\
     & learning rate & 0.1\\
     & \textit{max}\textsubscript{dep} & 2\\
     & \textit{max}\textsubscript{feat} & 5\\
      \bottomrule
    \end{tabular}
  \label{Tab3.0}
\end{table}
To compare the accuracy of ML models, the RMSE, MAE, and $\mathrm{R^2}$ score of ML models and the Arrhenius model have been calculated. 
It is clear to deduce that all ML models have fewer errors compared with the Arrhenius model reference (Figure \ref{fig:rmsemaer2}). This implies that ML models can accurately predict the diffusion coefficient and hence the introduction of ML provides an economical and time-saving route for predicting the diffusion coefficient of new systems rather than going through the whole process of conducting new MD simulations to obtain data, which requires heavy computational resources and efforts. 


\begin{figure}[H]
    \centering
    \includegraphics[scale=0.75]{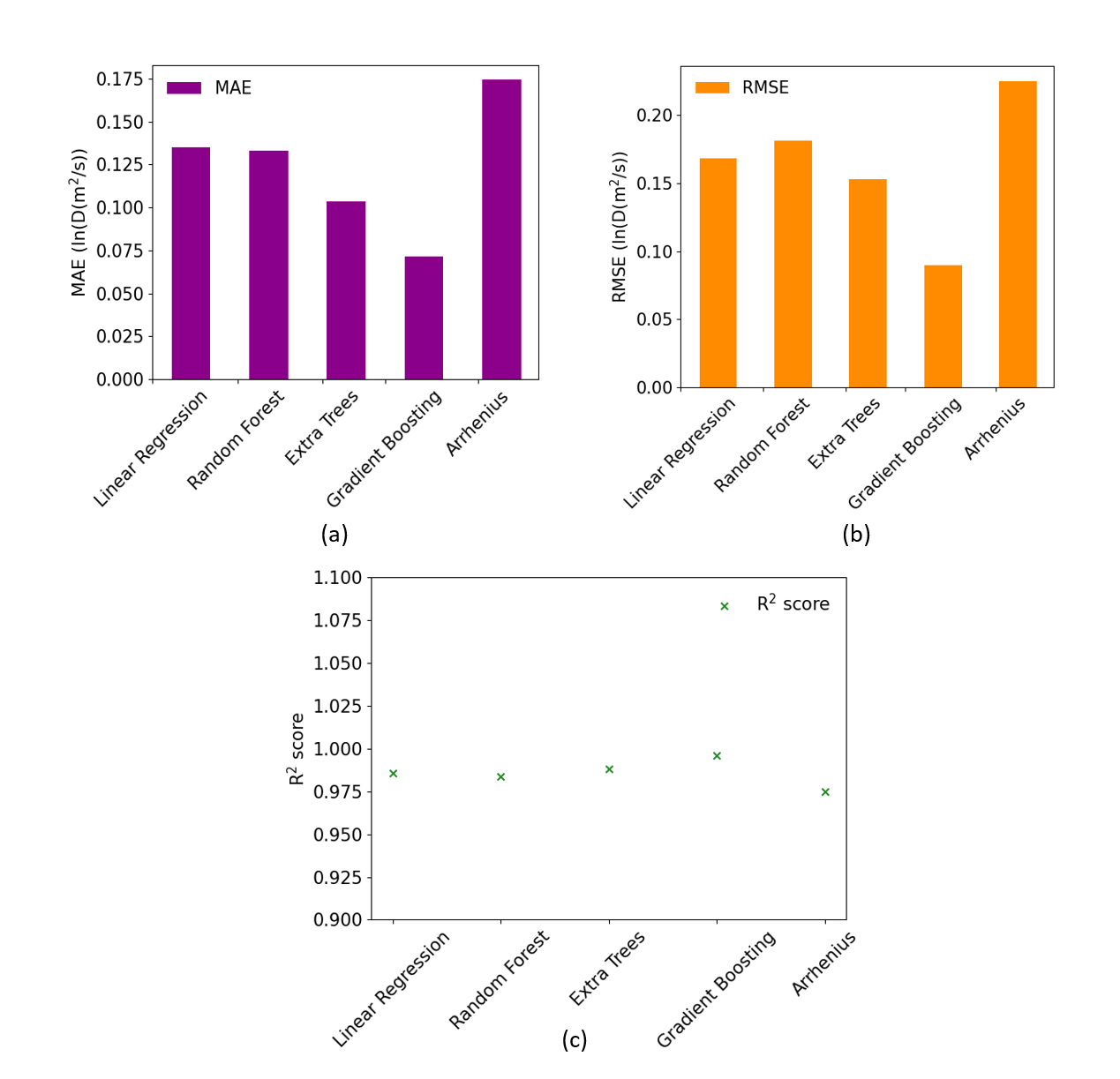}
    \caption{Metrics used to compare the accuracy of gradient boosting (GB), extra trees (ET), random forest (RF), linear regression (LR), elastic net (EN) and naive baseline (NB) methods which are calculated using the testing data set containing 85 entries. (a) Mean absolute error (MAE); (b) Root mean squared error (RMSE) and (c) $\mathrm{R^2}$ score.}
    \label{fig:rmsemaer2}
\end{figure}

Furthermore, we have compared $lnD$ from simulations with predicted $lnD$ for the GB and Arrhenius models in Figure \ref{fig:fvss}. It is observed that almost all points fall near the straight line with slope 1 from the GB model (Figure \ref{fig:fvss} (b)), whereas there are more divisions in the Arrhenius model (Figure \ref{fig:fvss} (b)). We therefore conclude that the machine learning based GB model accurately captures the fluctuation of diffusion coefficient of $\mathrm{H_2}$ with known parameters, which are generally considered in the realistic conditions. The GB model is then used to produce the feature importance rank (FIR), which is an additional measure of feature relevance. As a starting point, we calculate FIR that covers the entire temperature range (Table \ref{tab:fir} (a)). The temperature ($\sim$ 0.945) shows the strongest correlation with the diffusion coefficient of $\mathrm{H_2}$ among all features. The ranking of the correlation to the diffusion coefficient of $\mathrm{H_2}$ is $\mathrm{Ca^{+2}}$ ($\sim$ 0.032) $>$  $\mathrm{Na^+}$ ($\sim$ 0.019) $>$  $\mathrm{K^+}$ ($\sim$ 0.001). The value of $\mathrm{K^+}$ is so low that its effect on the diffusion coefficient becomes insignificant when combined with other ions as demonstrated in the previous observation (Figure {\ref{fig:triw} (b)}). The effect of pressure is insignificant ($\sim 0.002$). After that, we divided the data set into two parts in order to compute the FIR within low temperature range (298 K $\leq$ T $<$ 400 K, Table \ref{tab:fir} (b)) and high temperature range (400 K $\leq$ T $\leq$ 648 K, Table \ref{tab:fir} (c)). It is observed that the diffusion coefficient of $\mathrm{H_2}$ shows a strong dependence on the pressure (from $\sim$ 0.0005 to $\sim$ 0.007) as the temperature increases. This lends credence to our previous discussions in Figure \ref{fig:pe} regarding the pressure effect on the diffusion coefficient. Consequently, it is evident that the conclusions deduced in the previous discussions are in agreement with the results of FIR. 



\begin{figure}[H]
    \centering
    \includegraphics[scale=0.8]{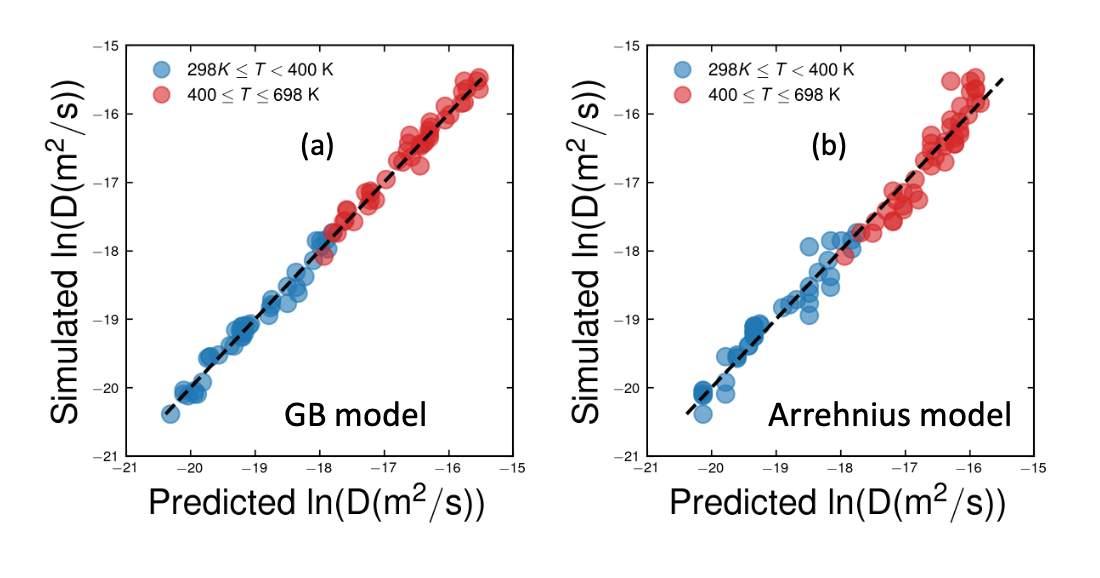}
    \caption{(a) Fitting data from gradient boosting; (b) fitting data from Arrhenius equation (Eq. \ref{eqn: straightfit}) versus data set from molecular dynamics (MD) simulations of ln(D ($\mathrm{m^2/s}$)) utilizing 197 training samples and 85 testing data samples.}
    \label{fig:fvss}
\end{figure}

\begin{table}
    \begin{subtable}[t]{0.9\textwidth}
    \centering
    \begin{tabular}{c|c}
    \toprule
       Feature  &  Importance value\\
    \midrule
        1/T & 0.9454 \\
        $\mathrm{c_{Ca^{2+}}}$ & 0.0322 \\
        $\mathrm{c_{Na{+}}}$ & 0.0192 \\
        P & 0.0021 \\
        $\mathrm{c_{K^{+}}}$ & 0.0011 \\
    \bottomrule
    \end{tabular}
    \caption{Overall data set containing samples with 298 K $\leq$ T $\leq$ 648 K.}
    \label{tab:firtrain}
    \end{subtable}
    \begin{subtable}[t]{0.4\textwidth}
    \centering
    \begin{tabular}{c|c}
    \toprule
       Feature  &  Importance value\\
    \midrule
        1/T & 0.7713 \\
        $\mathrm{c_{Ca^{2+}}}$ & 0.1711 \\
        $\mathrm{c_{Na{+}}}$ & 0.0550 \\
        $\mathrm{c_{K^{+}}}$ & 0.0021 \\
        P & 0.0005 \\
    \bottomrule
    \end{tabular}
    \caption{Data set containing samples with 298 K $\leq$ T $<$ 400 K.}
    \label{tab:firlow}
    \end{subtable}
    \begin{subtable}[t]{0.4\textwidth}
    \centering
    \begin{tabular}{c|c}
    \toprule
       Feature  &  Importance value\\
    \midrule
        1/T & 0.8167\\
        $\mathrm{c_{Ca^{2+}}}$ & 0.0916 \\
        $\mathrm{c_{Na{+}}}$ & 0.0812 \\
        P & 0.0073 \\
        $\mathrm{c_{K^{+}}}$ & 0.0032 \\
    \bottomrule
    \end{tabular}
    \caption{Data set containing samples with 400 K $\leq$ T $\leq$ 648 K.}
    \label{tab:firhigh}
    \end{subtable}
\caption{Feature Importance Rank (FIR) of five parameters estimated using gradient boosted method using the data set obtained from MD simulations.}
\label{tab:fir}
\end{table}
}

\section{Conclusion}{

We employ a computational framework of the combination of molecular dynamics (MD) simulation and machine learning (ML) models to determine the diffusion of hydrogen molecules in the extreme conditions present in saline aquifers. To the best of our knowledge, feeding the MD data into the ML-based prediction of diffusion coefficient of $\mathrm{H_2}$ by considering temperature, pressure, and ion concentrations ($\mathrm{NaCl}$, $\mathrm{Ca{Cl_2}}$, and $\mathrm{KCl}$) is reported for the first time. 
Initially, we have verified that there is no observable effect of the concentration of hydrogen on its diffusion coefficient in water. The diffusion coefficients of hydrogen and water molecules have been compared with the experimental data to confirm the validity of our simulation settings. A decrease in the diffusion coefficient of $\mathrm{H_2}$ with an increase in the cation concentration is evident. This implies that high salinity reservoirs are more suitable for the hydrogen storage due to the lower diffusion coefficients.

Furthermore, our results demonstrate that the temperature of systems is the major determinant of the diffusion coefficients of the $\mathrm{H_2}$, whereas the effect of pressure on the diffusion coefficient is inconsequential for low temperature conditions (298 K). The traditional quantitative description, namely, the Arrhenius formula, can well predict the temperature dependence of the diffusion coefficient of $\mathrm{H_2}$ in brine at the constant pressure. However, we find that the diffusion coefficient of $\mathrm{H_2}$ in brine with low ion concentrations reduces significantly with the pressure at 648 K, implying the inaccuracy of the Arrhenius formula considering the effect of pressure and compositions of individual cations. The models which can incorporate the versatile behaviour of the diffusion coefficient, such as Zhao's model\cite{Zhao2021}, requires the information of viscosity and density of the solution, which are indirect parameters requiring further experiments or simulations.

Therefore, we have applied four ML algorithms (linear regression (LR), random forest (RF), extra trees (ET), and gradient boosting (GB)) to obtain the best model considering the accuracy, completeness, and simplicity. All the models show higher accuracy than the Arrhenius model from the evaluation of root mean squared error (RMSE), mean absolute error (MAE), and $\mathrm{R^2}$ score. When compared to other models, the GB model is the one that provides the best accurate predictions regarding the diffusion coefficients of hydrogen.
In addition, the feature importance rank deduced using the GB algorithm is in agreement with the conclusions obtained about the effect of different parameters on the diffusion coefficient. 

Overall, our work demonstrates that the Arrhenius fitting model can predict the diffusion coefficient for the low temperature regime (T$\leq$400 K). However, for a wide range of temperature and pressure range  in saline aquifers, using MD simulations to feed GB model is a promising approach to define a model which predicts the diffusion behavior of $\mathrm{H_2}$ and we believe that this computational framework has a significant potential for advancing the development of the large-scale hydrogen storage over seasonal time scales. }
}
\bibliography{citations,ML,intro}
\end{document}


\pagebreak

\subsection{Validation}

\begin{table}[H]
\centering
\caption{The diffusion coefficient and viscosity data from experiments\cite{Wise1966, Korson1969} at different temperatures ($P$ = 1 atm) used for the fitting of Wilke-Chang equation:}
\label{table:2}
    \begin{tabular}{c|c|c}
    \toprule
        T (K) & $\mu$ (cP) & D $\mathrm{(10^{-8} m^2/s)}$ \\
        \midrule
        283	& 1.3069 & 0.46 \\
        293	& 1.0020 & 0.5 \\
        303	& 0.7975 & 0.7 \\
        313	& 0.6532 & 0.83 \\
        323	& 0.5471 & 0.97 \\
        333	& 0.4666 & 1.31 \\
    \bottomrule
    \end{tabular}
\end{table}

\begin{figure}[H]
	\centering
	  \includegraphics[scale=0.7]{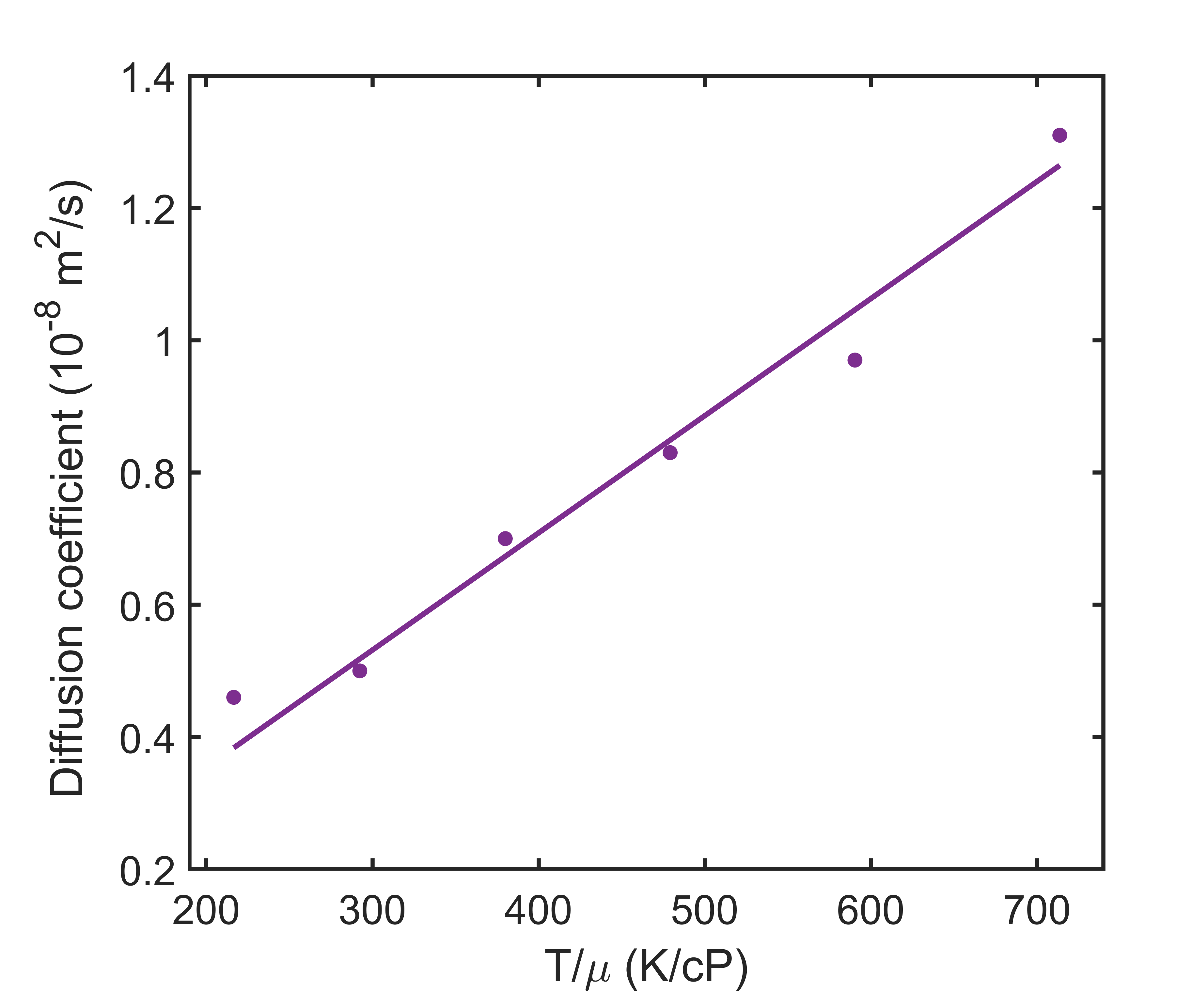}
	  \caption{Wilke-Chang fit of the variation of the diffusion coefficient of $\mathrm{H_2}$ in pure water versus $\mathrm{T/\mu}$ for the data in Table \ref{table:2} at $P$ = 1 atm.}
	  \label{fig:wilke}
\end{figure}

From Figure \ref{fig:wilke}, we can infer that the $\mathrm{H_2}$ + $\mathrm{H_2O}$ system obey the Wilke-Chang equation. From this fit, D = 0.593x$\mathrm{10^{-8}}$ $\mathrm{m^2/s}$ at $T$ = 298 K and $P$ = 1 atm.

\subsection{Interaction analysis}

\begin{figure}[H]
	\centering
	  \includegraphics[scale=0.85]{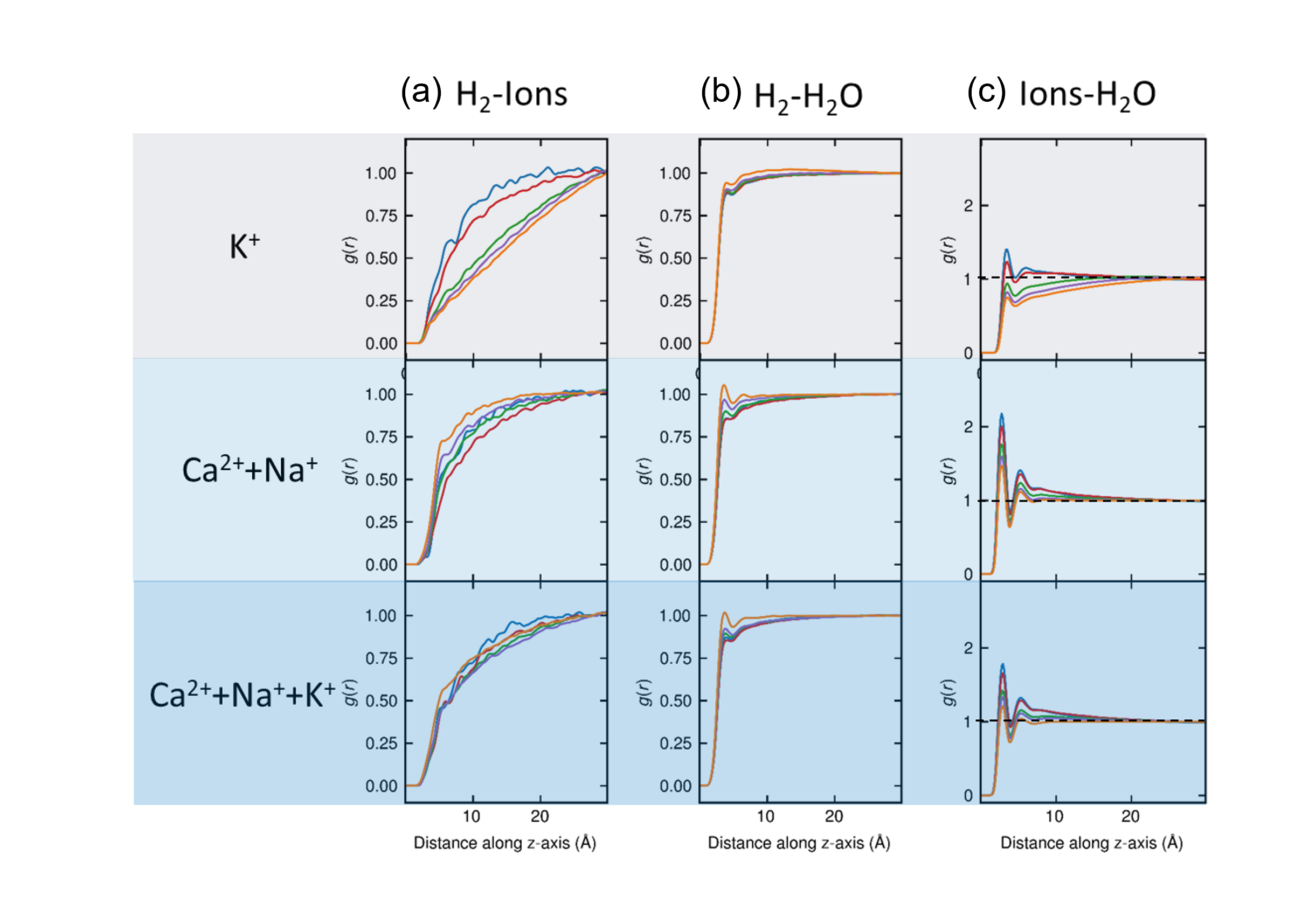}
	  \caption{The radial distribution function $g(r)$ (a) between $\mathrm{H_2}$ and cations, (b) between $\mathrm{H_2}$ and cations, and (c) between $\mathrm{H_2}$ and cations. Single cation brine system $\mathrm{K^+}$, binary cation brine system $\mathrm{Na^+}+\mathrm{Ca^+}$, and ternary cation brine system $\mathrm{K^+}+\mathrm{Na^+}+\mathrm{Ca^+}$ are shown from the top to bottom panel. Temperature and pressure of all systems are at $T$ = 648 K and $P$ = 218 atm, respectively.}
	  \label{fig:rdf}
\end{figure}

The intermolecular radial distribution function $g(r)$ is defined
as the probability of finding the specific cation at a distance $r$ from a given $\mathrm{H_2}$ molecule (center of mass). Single cation systems containing $\mathrm{K^+}$, binary cation systems containing $\mathrm{Na^+-Ca^{2+}}$, ternary cations containing $\mathrm{Na^+-Ca^{2+}-K^+}$ have been chosen to calculate the intermolecular radial distribution function at the condition of high temperature and pressure ($T$ = 648 K and $P$ = 218 atm) in Figure \ref{fig:rdf}. We find that $g(r)$ of systems with larger molality is evidently higher than that with smaller molality for $\mathrm{K^+}$ (Figure \ref{fig:rdf} (a)) cases. This indicates that the interaction between $\mathrm{H_2}$ and cations decreases by adding more ions in the system. Lower interaction between $\mathrm{H_2}$ and the cation causes lower aberration to the movement of $\mathrm{H_2}$. Especially, the $g(r)$ between $\mathrm{H_2}$ and $\mathrm{H_2O}$ in systems containing single $\mathrm{K^+}$ ion is lower than that containing other ions. As the physical significance of diffusion coefficient correlates it with the movement of molecules\cite{Murphy2015}, it can be concluded that the disturbance in the motion of molecules in KCl systems due to decreased interactions can cause the diffusion coefficient to increase. Moreover, we find that all radial distribution functions of ternary containing systems are more similar to the binary cation systems containing $\mathrm{Na^+-Ca^{2+}}$ rather than single cation systems containing $\mathrm{K^+}$. This indicates that the structure-making ions show the dominant effect in the ternary cations systems.

\subsection{Machine Learning}

The equation of the (i) linear regression (LR) model to a data set (n features and m samples) is 
\begin{equation}
Y = \beta_{0} + X\beta
\label{eq1}
\end{equation}
where $X$ is the rectangular input matrix (dimension of \textit{m}$\times$\textit{n}) containing the features data $X_{i1}$, $X_{i2}$, $X_{i3}$, $X_{i4}$ and $X_{i5}$ which are T, P, $\mathrm{c_{Na^+}}$, $\mathrm{c_{Ca^{2+}}}$ and $\mathrm{c_{K^+}}$ of the i'th sample, respectively; $Y$ is the corresponding output vector containing the diffusion coefficient data (dimension of $m$$\times$1) in which $Y_i$ is the diffusion coefficient of the i'th sample;  $\beta_0$ is the intercept vector (dimension of $m$$\times$1) with identical elements and $\beta$ is the coefficient vector (dimension of $n$$\times$1) in which $\beta_j$ is the coefficient of the j'th feature.
Scikit-learn applies the LR model with the concept of "ordinary least square" (OLS). The given data set is fitted with vectors $\beta_0$ and $\beta$ to minimize the sum of squared residuals ($SSR$), which is mathematically interpreted as follows:
\begin{equation}
SSR=\min_\beta {||A||_{2}}^2 = \min_\beta {||(\beta_{0} + X\beta) - y||_{2}}^2
\label{eq2}
\end{equation}
where \textit{y} is the actual value matrix from the data set; ($\beta_{0} + X\beta$) is the predicted value matrix and the symbol $||$ $||_2$ is the modulus of root of sum of squared elements of the column matrix.



\begin{figure}
    \centering
    \includegraphics[scale=0.43]{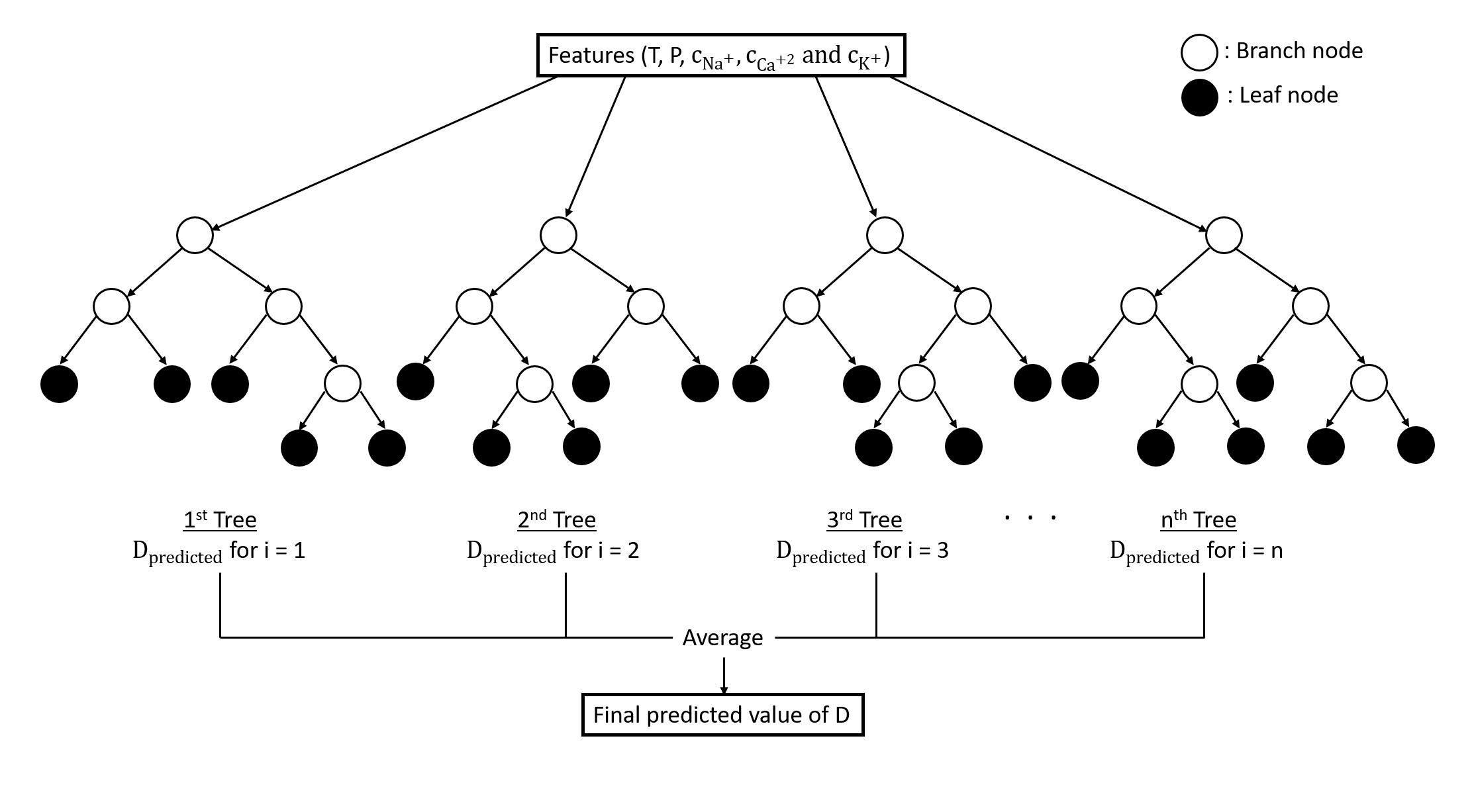}
    \caption{Schematic diagram of the working of RF and ET models (four branches are depicted for clarity) to give predictions of the diffusion coefficient. The data set contains 5 features: Temperature (T), pressure (P), and three cation concentrations ($\mathrm{c_{Na^+}}$, $\mathrm{c_{Ca^{2+}}}$ and $\mathrm{c_{K^+}}$). }
    \label{fig:RF}
\end{figure}

In addition to the linear regression model, three other regression models based on ensemble methods are also used for data prediction. Ensemble methods combine estimators which are base machine learning (ML) models to obtain an optimal predictive model. The (ii) random forest (RF) and (iii) extra trees (ET) models belong to the averaging method that contains a certain number of decision trees (estimators) \cite{Ensemble}. Consequently, these decision trees make their predictions independently, which will be averaged to give an overall prediction of the given data set as shown in Figure \ref{fig:RF}. Each decision tree splits the MD data set into subgroups called nodes. At the beginning, the parent node which is the starting node, will be split into two children nodes. The children nodes then continue being split into further nodes. If a node doesn't possess any children node, this is called a leaf node. The data is run through all the trained decision trees until they reach the respective leaf nodes which consist the prediction for the specific sample. The differences between RF and ET models are:  (a) the RF model uses bootstrap samples which are sub samples of the input data with replacement\cite{Ensemble}, whereas the ET model uses the whole input data set; (b) while splitting a node, the RF model uses the best feature whereas the ET model uses a random feature among a random subset of features of size $\mathrm{\sqrt{n}}$. The hyperparameters tuned for RF and ET models are the number of estimators, maximum number of features and the maximum tree depth, which signifies the number of branches of each tree. Firstly, deciding on the number of estimators and maximum number of features can save computer time as there is no significant improvement in the model's accuracy after a certain value. Secondly, the maximum depth of the tree can be controlled to prevent the model from overfitting the training data set.

The (iv) gradient boosting (GB) model belongs to the boosting approach\cite{Ensemble} in which the estimators are built sequentially by taking into account the previous estimator's results. The estimators are decision trees in case of GB model. Initially, the average of the diffusion coefficients from the training data set ($\mathrm{y_{avg}}$) is calculated. The psuedo residuals which is the difference between the diffusion coefficients from the training data set and the average value are calculated and then used to train the first decision tree. The psuedo residual obtained from the leaf node of the decision tree ($\mathrm{PR_{i1}}$ where i is i'th sample and 1 denotes the first decision tree) is added to the average diffusion coefficient with a learning rate ($\mathrm{\alpha}$) to obtain the predicted diffusion coefficient from the first decision tree ($\mathrm{y_{pi1} = y_{avg} + \alpha \cdot PR_{i1}}$). After calculating all the diffusion coefficients using the first decision tree, new psuedo residuals are calculated which is the difference between diffusion coefficients from the training data set and diffusion coefficients from the first decision tree. New psuedo residuals are lower than the initial psuedo residuals. The second decision tree is trained using these new psuedo residuals and the psuedo residual obtained from the leaf node of this tree ($\mathrm{PR_{i2}}$) is added to the prediction from the first decision tree to get the prediction from the second decision tree ($\mathrm{y_{pi2} = y_{pi1} + \alpha \cdot PR_{i2}}$). Hence, decision trees are added sequentially with a learning rate to reduce errors (psuedo residuals) from previous decision trees. This trained sequence of decision trees is used to predict the diffusion coefficient for any new sample. The hyperparameters tuned here are the learning rate, number of estimators, maximum tree depth and maximum number of features. The naive gradient boosting is the algorithm with a learning rate as 1. Learning rate and the number of estimators are correlated. Suppose the learning rate is less than 1, fewer corrections are made for each estimator, thus more estimators are required. If the learning rate is greater than 1, more corrections are made per estimator, so fewer estimators are required.

\begin{figure}
    \centering
    \includegraphics[scale = 0.6]{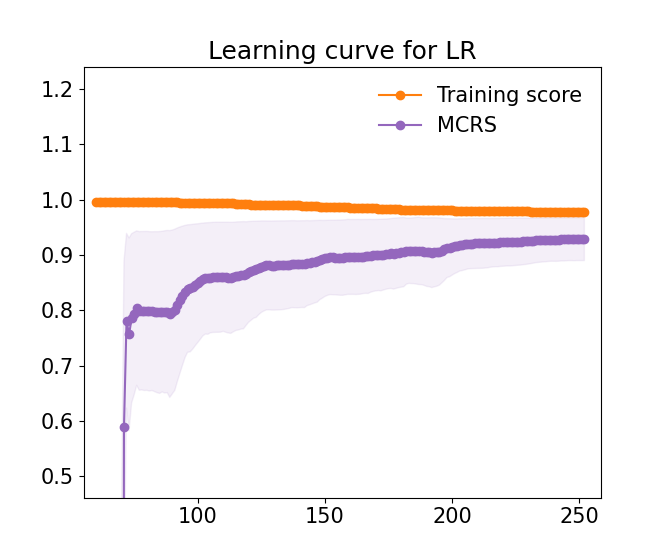}
    \caption{Learning curve obtained using linear regression (LR) where MCRS is the mean cross-validated $\mathrm{R^2}$ score calculated over 10 cross-validation splits of the data set and the training score is the $\mathrm{R^2}$ score calculated using training samples.}
    \label{fig:learncurve}
\end{figure}

To manifest the split of the data set into training and testing samples, the learning curve which is the curve of the prediction accuracy vs the number of training samples is obtained using LR is studied. The learning curve as shown in Figure \ref{fig:learncurve} suggests that the prediction accuracy which is the mean cross-validated $\mathrm{R^2}$ score calculated over 10 cross-validation splits of the data set does not vary significantly beyond 197 training samples marking the train - test split to be 70\%. 

\begin{figure}[H]
    \centering
    \includegraphics[scale=0.92]{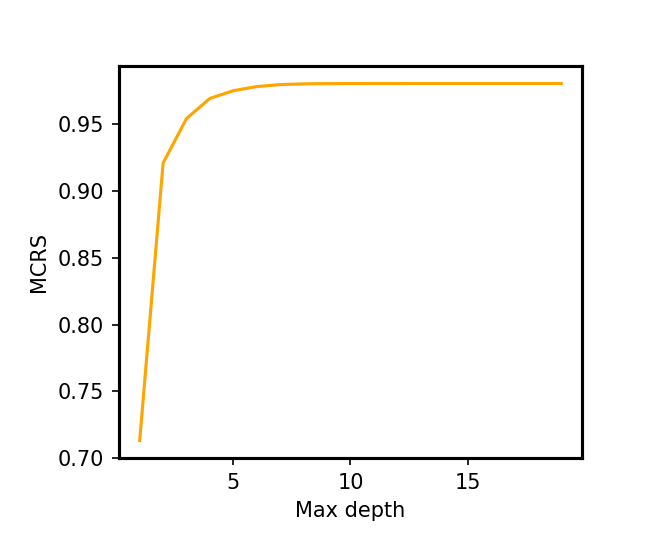}
    \caption{The variation of mean cross-validated $\mathrm{R^2}$ score with respect to the maximum depth of the trees for the random forest model.}
    \label{fig:mcrsRF}
\end{figure}

\begin{figure}[H]
    \centering
    \includegraphics[scale=0.73]{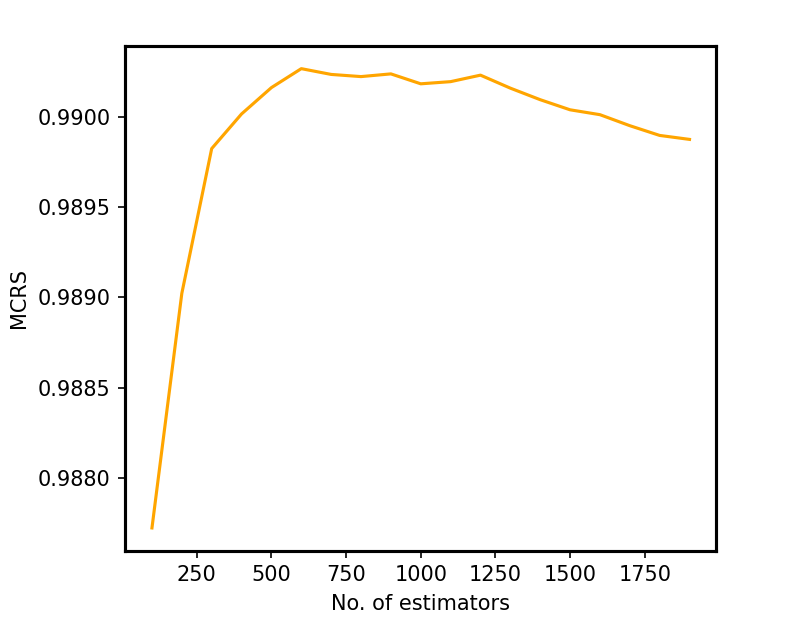}
    \caption{The variation of mean cross-validated $\mathrm{R^2}$ score with respect to the number of estimators for a learning rate = 0.1, maximum tree depth = 2 and maximum number of features = 5 for the gradient boosting model.}
    \label{fig:mcrsGB}
\end{figure}

The optimization of hyperparameters for each model is characterized from the MCRS calculated over ten cross-validation splits. Firstly, we find that the accuracy of RF and ET models saturates as the maximum depth of trees increases of 8 as shown in Figure \ref{fig:mcrsRF}. Usually, no restriction on the growth of the trees may cause over-fitting, but in our case, any efforts to restrict the growth caused the MCRS to decrease. Our findings suggest that a maximum depth of 13 for RF and 10 for ET provides the best accuracy.
Additionally, the optimum number of estimators and the maximum number of features has been characterized as 800 and 5; and 100 and 5 for RF and ET models, respectively. Any attempt to reduce the maximum number of features reduced the accuracy of the models substantially. 
Secondly, examining the GB model, we find that a low magnitude of the learning rate is required to obtain good accuracy as the number of estimators increases as shown in Figure \ref{fig:mcrsGB}. The optimum learning rate and the number of estimators are calculated as 0.1 and 600 for the best accuracy in this work, respectively. The optimum maximum tree depth is found to be 2. Reducing the maximum number of features caused the accuracy of the model to reduce significantly and hence a value of 5 provides the best result.


\bibliography{citations}